# Boltzmann Physical Kinetic for Deep Bed Filtration:
# Homogenization by solving operator equation in Hilbert space

Dinariev, O.Yu.[1], Pessoa Rego, L. A.[2], Bedrikovetsky, P.[2]

[1]Institute of the Earth Physics, Russian Academy of Sciences; [2]University of Adelaide, Australia

**Abstract**

This paper develops a modified version of the Boltzmann's equation for micro-scale particulate flow with capture and diffusion that describes the colloidal-suspension-nano transport in porous media. An equivalent sink term is introduced into the kinetic equation instead of non-zero initial data, resulting in the solution of an operator equation in the Fourier space and an exact homogenization. The upper scale equation is obtained in closed form together with explicit formulae for the large-scale model coefficients in terms of the micro-scale parameters. The upscaling reveals the delay in particle transport if compared with the carrier water velocity, which is a collective effect of the particle capture and diffusion. The derived governing equation generalizes the current models for suspension-colloidal-nano transport in porous media.

**Nomenclature**

*Latin letters*

$A$ – area of cross-section, $L^2$

$c$ – suspended particle concentration, $L^{-3}$

$c_0$ - initial suspended particle concentration, $L^{-3}$

$C_v$ – coefficient of variation

$d$ – site occupation function $d(c)$,

$D_\mu$ - *molecular diffusion*

$F$ – individual suspension function for particles with size $r$, $L^{-5}$

$f$ – particle concentration density function, $TL^{-4}$

$H$ – Hilbert space

$l$ – mixing (correlation) length, $L$

$L$ – core length, $L$

$p$ – pressure, $MT^{-2}L^{-1}$

$P$ – projection operator in Hilbert space

$Pe$ – Peclet number

$q$ – particle flux, $L^{-2}T^{-1}$

$r$ – pore radius, $L$

$r_s$ – particle radius, $L$

$R_{ij}$ – transport coefficients

$s$ – source term in Boltzmann's equation

$t$ – time, $T$

*T – dimensionless time, PVI*

$U$ – Darcy's velocity of the carrier fluid, $LT^{-1}$

$v$ – velocity, $LT^{-1}$

$x$ –Cartesian coordinate, $L$

$X$ - dimensionless Cartesian coordinate

*Greek letters*

$\alpha_L$ – dispersivity coefficient, L

$\varepsilon$ - capture rate, $L^{-3}T^{-1}$

$\Theta$ – delay number

$\lambda$ – filtration coefficient, $L^{-1}$

$\mu$ – viscosity, $ML^{-1}T^{-1}$

$\tau$ – diffusion relaxation time, $T$

$\phi$ – porosity

$\psi_0$ – velocity distribution function, $L^{-1}T$

$\psi_1$ – normalized velocity distribution function

$\omega$ – probability of particle capture by the sieve

$\Omega$ – capture rate coefficient

*Abbreviations*

*BTC – breakthrough curve*

*PVI – pore volume injected*

## 1. Introduction & Motivation

*Suspension-colloidal-nano flows in porous media* occur in numerous environmental, agricultural and water-management processes and technologies. The incomplete list encompasses flows near artesian injection and production wells, propagation of viruses, bacteria and other mechanisms of aquifer contamination, nanotechnologies to fix the contaminants in the rock, fresh-water storage in aquifers, vadoze zone dynamics, fines migration in subterranean formations, and suspension invasion during well drilling, injection and produced water disposal in oilfields [11, 12, 19, 20, 21, 32, 54, 60, 61]. The main features of the above mentioned processes are particulate transport and particle capture by the rock.

Figure 1a exhibits some of the pore scale mechanisms of particle retention: size exclusion, electrostatic attachment, straining, bridging, diffusion into dead-end pores, and gravity segregation [11, 32]. Figure 1b shows the velocity distribution in a reference porous-media volume. Other distributed parameters at the pore- and core scales are surface roughness, heterogeneity of the electrostatic surface charges and zeta-potentials due to multiple minerals composing the rock, and the particle and pore sizes and shapes [19, 32, 58]. Micro-scale distributions of physical parameters for colloidal and nano flows in porous media yield stochastic transport equations [21, 22, 30, 31]. Derivation of effective upscaled equations accounting for the micro-scale parameter distributions is essential for modelling of the above-mentioned natural processes and industrial technologies [28, 34, 39].

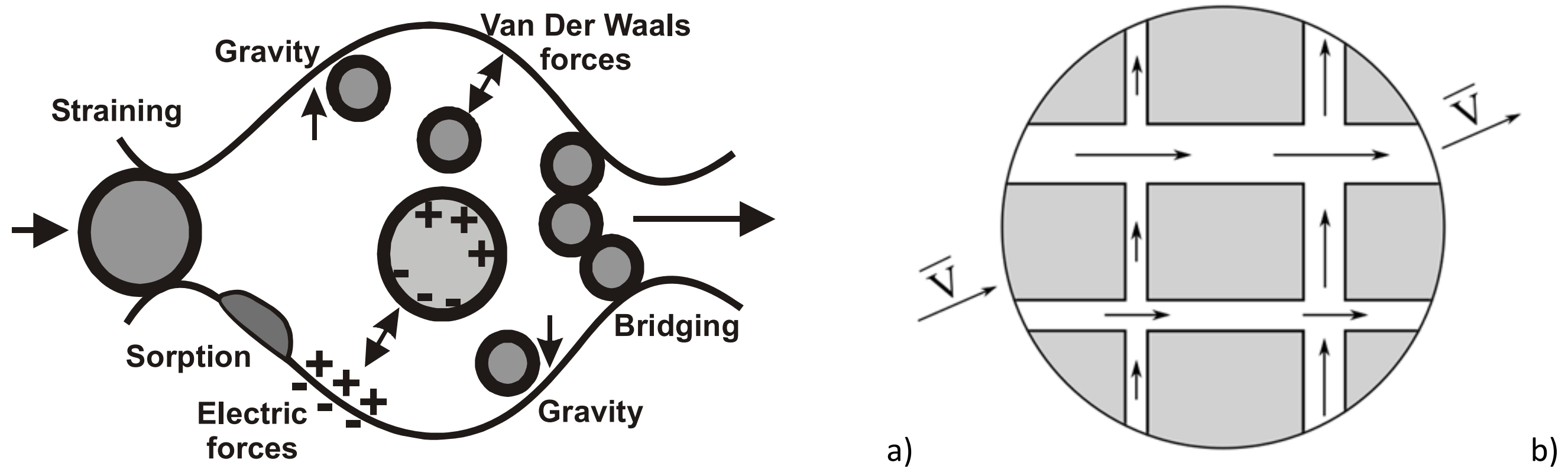

Figure 1. Stochastic physics factors of particle capture: a) Different capture mechanisms; b) Distribution of fluid velocity at the pore-network scale.

*The traditional large-scale mathematical model for suspension-colloidal-nano transport* includes a mass balance equation for suspended and retained particles [4, 5, 36]

$$\phi\frac{\partial c}{\partial t}+\frac{\partial q}{\partial x}=-\varepsilon\,, \tag{1}$$

where $\phi$ is the porosity, $c$ is the concentration of suspended particles, and $q$ is the overall particle flux. The suspension concentration $c$ is equal to the number of particles per unit of the liquid volume.

The total particle flux $q$ consists of advective and diffusive/dispersive components

$$q=vc-D\frac{\partial c}{\partial x}\;, \tag{2}$$

$$D=D_{\mu}+\alpha_L\left|v\right|, \tag{3}$$

where $v$ is the carrier water velocity, $D_{\mu}$ is the molecular diffusion coefficient, $\alpha_L$ is the dispersivity coefficient, and $|\;|$ corresponds to absolute value (modulus) [4, 51].

System (1-3) is closed by the equation for the particle retention rate. Traditionally it is assumed that the rate is proportional to the modulus of the advective flux [36, 38, 63-66]

$$\varepsilon=\lambda c\left|v\right|. \tag{4}$$

The proportionality coefficient $\lambda$ in Eq. (4) is a particle capture probability per unit length of the particle trajectory. The coefficient $\lambda$ is called the filtration coefficient. The retention-rate expression (4) is used in vast majority of the models for suspension-colloidal-nano transport in porous media [11, 12, 56, 63].

The modification of the classical model (4) assumes that the kinetic rate is proportional to the overall particle flux [1, 66]:

$$\varepsilon=\lambda\left|q\right|. \tag{5}$$

*Shortcomings of the models (4) and (5)* Equation (5) describes a particle capture by the vacancy independently of whether it is advection or dispersion/diffusion that brought the particle to the vicinity of a retention site (vacancy), while Eq. (4) does not account for particles brought to the vacant site by diffusion/dispersion. As it follows from the decomposition of the particle trajectory into the advective displacement and diffusive jumps, Eq. (5) calculates the capture rate for the particles jumping forward proportionally to the trajectory length, while the capture of particles that jump back is calculated by Eq. (5) twice [66]. Therefore, modification of Eq. (5) is required.

Eq. (5) was used in the governing system for suspension-colloidal-nano flows in [20, 62, 66]. Works [41, 44] derived Eq. (5) by averaging the micro-scale continuous Markov chains (Fokker-Plank equations). Moreover, papers [17,18] validated the model (1-3, 5) by comparison with laboratory corefloods.

In the case of particle diffusion in stagnant water ($v$=0), where particles are captured by the rock during Brownian jumps, an initially uniform concentration profile declines with time and remains uniform across the rock. In this case, Eq. (4) yields capture-free molecular diffusion; this is a shortcoming of the model (4). Both terms in the flux expression (2) are zero for the case of uniform concentration profiles, which also results in zero particle capture by Eq. (5) and is a shortcoming of this model.

Consider suspension injection into a clean bed. In this case, the concentration gradient is negative. Substituting Eqs. (3) and (5) into mass balance (1) yields a delay in particle speed if compared with the carrier water velocity from $v$ to $v-\lambda D$. For large filtration coefficients $\lambda>v/D$, the particle speed becomes negative resulting in the particle counter flow; this is also an unexplained shortcoming of the model (5).

*State-of-the-art* Currently, core-scale mathematical models for suspension-colloidal-nano transport in porous media closely match the laboratory data in the vast majority of experimental studies and are successfully used for large-scale laboratory-based reservoir behavior predictions [8, 9, 11, 12, 19]. However, poor match of the laboratory data observed in numerous specific cases (hyper-exponential and non-monotonic retention profiles, significant deviations during simultaneous matching the breakthrough concentrations and retention profiles) has resulted in modifications being made to the macro scale transport systems via derivations from micro-scale equations [38, 45, 64, 65]. Besides, the methods for determination of the pore-scale properties were significantly developed recently, allowing for detailed model validation [8, 53]. Despite a long history, *the area of upscaling / homogenization / averaging for microscale colloidal transport in porous media is still not fully explored.*

During the last three decades, the upscaling in colloidal, solute and reactive transport in porous media with stochastic heterogeneity has been significantly developed [28, 29, 49, 53, 59]. This includes the cases of closed upscaled systems as well as hybrid multi-scale numerical models [13, 22, 23, 30, 48]. Rock heterogeneity is established as a result of numerous geological processes; the correlation length of the heterogeneity is an important reservoir parameter [10, 21, 34]. Yet, often the correlation length is significantly smaller than the size of the boundary problem (the reservoir). In this work we discuss this important particular case, where the correlation length (system dispersivity) is equal to zero. In other words, the reservoir stochastic properties are uncorrelated.

Different generalizations of systems (1-4) and (1-3, 5) have been obtained by averaging/upscaling of stochastic micro-scale models. The stochastic models include trajectory analysis [50], continuous random walk models [57], random filtration coefficient [32, 63, 64], and population balance systems [5-7, 58]. The micro-scale modelling results in significant enrichment of the hydrodynamic equations (1-5) for deep bed filtration. The random-walk modelling leads to time and space memory [57]. Asymptotic averaging of the continuous random-walk Master equation with distributed times and lengths of the jumps and fixed capture probability yields scalar velocity-independent capture rate and adds timely and mixed diffusion terms in the mass balance equation (1), which become elliptic [56, 57].

Homogenization of diffusion-free micro-scale population-balance equations leads to either system (1-4) or (1-3, 5) without diffusion. Exact upscaling of the population balance system for mono-size suspension in micro-heterogeneous porous media yields the extended system (1-4) that contains the retention-dependent filtration function and accounts for pore-space accessibility for finite-size particles (Bedrikovetsky 2008). The upscaling procedure is reversible, allowing for restoration of micro-scale behavior from the upper-scale measurements, i.e. for downscaling [6]. Low-concentration multi-size populations perform independent deep bed filtration in porous media with stochastic micro-scale heterogeneity, i.e. equations for transport of different populations turn out to be independent; so the system does not allow upscaling.

Upscaling of size-distributed suspension-colloidal-nano transport with particle attachment and no diffusion results in changing the suspended concentration $c$ in Eq. (4) to non-linear suspension function $f(c)$ and adds a third independent equation for kinetics of the rock surface occupation by the attaching particles [7]. Upscaling of population-balance equations for colloidal-nano transport in porous media accounting for diffusion/dispersion *is not available.*

Boltzmann's modelling of homogeneous fluid transport in porous media was achieved by the so-called dusty gas model, where the "immobile" porous matrix was represented by "heavy" particles, resulting in a binary particle flow [46]. The kinetics equation and the asymptotic Chapman–Enskog method [14, 15, 43] allow the derivation of hydrodynamics equations for flow and diffusion of gases in porous media. Darcy's law is obtained as an asymptotic case of heavy particles. The model is valid for rarefied gases. It *describes molecular but not convective diffusion*.

Different formulations of Boltzmann's transport system in periodical and quasi periodical media and in bounded channels result in hybrid numerical algorithms with large-scale dependency of the micro-scale behavior [3, 16, 23, 33, 35, 37, 42].

Another approach for the kinetic theory of flows in porous media was undertaken in [55], where the porous medium was represented by a surface that is present in any reference volume of the rock. The expression for the particle collision integral with the surface was derived. The averaged transport coefficients admit a transparent physical interpretation, which coincides with that obtained by fluctuation theory. The above-mentioned kinetics equations have been derived and upscaled for transport of homogeneous fluid; the similar approach for particulate transport by a carrier fluid *has not been pursued*.

*In the present paper*, Boltzmann's kinetic equation is used to describe the evolution of particle velocity distribution during deep bed filtration. The mixing (correlation) length of the porous media is used as a constant property of the rock to describe the relaxation, instead of a constant relaxation time; it leads to proportionality of all transport coefficients to the flow velocity. The introduction of a sink term into Boltzmann's equation instead of non-zero Cauchy data (so called sink-source method, developed in works [24-27]) was applied to the linear kinetics equation. The decomposition of the Fourier transform of Boltzmann's operator in two orthogonal subspaces of Hilbert space allows exact averaging of the flux and explicit form of the hydrodynamic equations and constitutive relationships. The obtained averaged model on the large-scale significantly differs from the classical deep bed filtration equation and its later modifications.

The structure of the text is as follows. Section two formulates Boltzmann's kinetic equation for flow of particles with stochastically distributed velocities and introduces the sink-source method. Section three represents the Boltzmann's equation in Hilbert space and performs the averaging, leading to explicit constitutive relations and a closed macro scale system. Section four analyses the upscaled large-scale model and properties of the model coefficients. Section five analyses three particular cases of colloidal-suspension transport in the framework of the derived averaged transport equation. Section six discusses the model validity and limitations. Section seven concludes the paper.

## 2. Boltzmann's micro-scale equation for suspension-colloidal-nano transport in porous media

In this section we introduce Boltzmann's equation for flow of suspended or colloidal particles in porous media with the main assumptions (section 2.1). The so-called sink-source method [24-27] substitutes the Cauchy problem with non-zero initial data by the sink-source term in Boltzmann's mass balance equation. The conservation law is formulated for the averaged mass balance equation (section 2.2).

### 2.1. Assumptions of the Boltzmann's kinetics model

*Velocity-distribution of colloid* Let us consider the kinetic description of particle-colloid transport in one-dimensional (1D) liquid flow, where the particle velocities are randomly distributed. The state of the suspension at any arbitrary moment is determined by the particle concentration density function [14, 15, 43]

$$f = f\,(t, x, v)\,, \tag{6}$$

which describes the particle velocity distribution (Figure 1b). The non-negative function $f(t,x,v)$ describes the time evolution of the distribution of particles which move with velocity $v \in R^{dv}$, in the position $x \in R^{dx}$ at time $t>0$. Further we discuss 1D transport $d_v = d_x = 1$.

Here $v$ is the interstitial velocity in the pore where the particles move, $f\Delta x\Delta v$ is the number of particles in the rock volume $\Delta x$ with velocities varying in the interval $\Delta v$. The inhomogeneous velocity distribution is determined by the pore space geometry and particle diffusion.

By definition, the density of particles $c=c(t,x)$ and the particle flux $q=q(t,x)$ can be calculated from the distribution function (6)

$$c(t,x)=\int f(t,x,v)\,dv, \tag{7}$$

$$q(t,x)=\int v f(t,x,v)\,dv\,. \tag{8}$$

We assume that the density function (6) can undergo changes due to the following effects: transport in the porous space, capture of particles and influence of the liquid, collisions with other particles and pore walls on the ensemble of particles.

We assume that during a steady-state flow with no capture, the particle velocity distribution is also steady-state, given by a density function $\psi_0= \psi_0(v)$. Here $\psi_0(v)>0$, and the integral of $\psi_0(v)$ in $v$ from minus infinity to infinity is equal to one.

While the capture process changes the particle velocity distribution, the interaction with the liquid and the matrix tends to make the velocity distribution close to the fixed distribution function $\psi_0= \psi_0(v)$. We assume that $\psi_0(v)$ is established during the capture-free steady-state flow of a suspension through a porous medium with some delay.

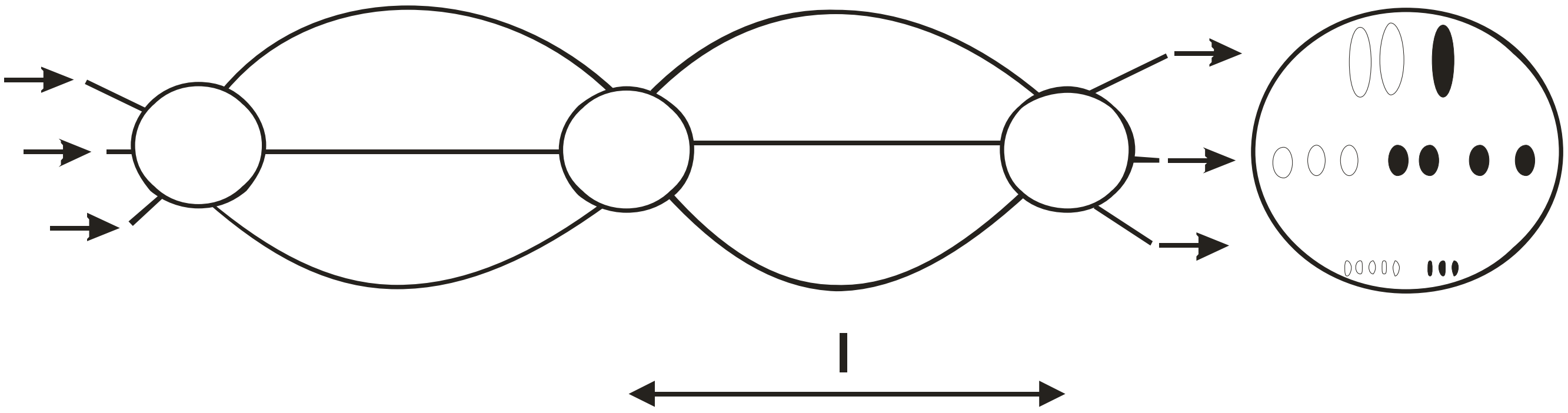

a) b)

Figure 2. Geometric model of porous media - bundle of parallel capillary alternated by mixing chambers: a) “vertical” cross section of the rock in plane *(x,z)*; b) cross section of the model-rock in plane *(y,z)*.

*Example.* Let us discuss the geometric porous-media model as a set of parallel capillary alternated by mixing chambers (Figure 2). The particles move in the pores that are larger than themselves, mix up in the chambers, and are captured at the exit of chambers in smaller pore throats and inside the pores (Figure 3).

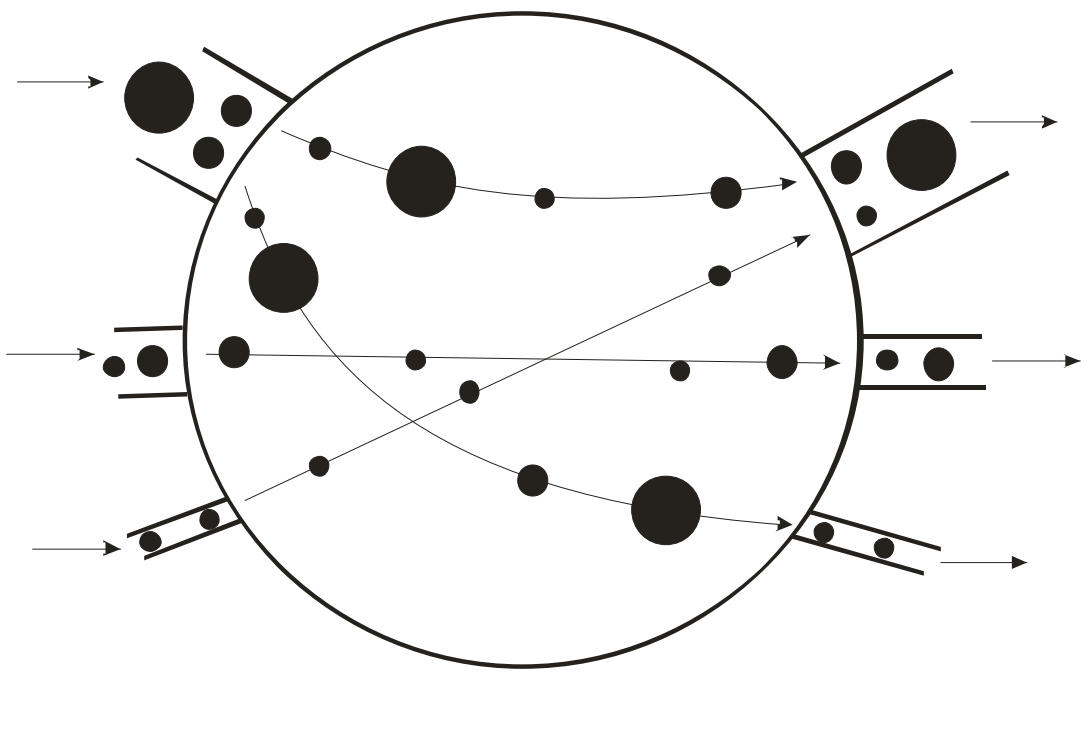

Figure 3. Schematic of complete suspension mixing in a single chamber

The assumption of Poiseuille’s velocity profile in each pore links the averaged particle velocity to the pore radius by

$$v = -\frac{r_p^2}{8\mu}\frac{\partial p}{\partial x}. \tag{9}$$

The mixing distributes the particle velocities according to the pore size distribution: the probability that the particle velocity falls between $v$ and $v+dv$ is equal to the probability that the particle size is between $r_p$ and $r_p+dr_p$:

$$\psi_0(v)dv = g(r_p)dr_p, \tag{10}$$

where $g(r_p)$ is the pore size distribution.

It follows from Eqs. (9, 10) that

$$\psi_0(v) = g(r_p)\frac{dr_p}{dv} = \frac{g\left(\sqrt{v/\eta}\right)}{2\sqrt{\eta v}}, \tag{11}$$

where

$$v = \eta r_p^2, \eta = -\frac{1}{8\mu}\frac{\partial p}{\partial x} \quad . \tag{12}$$

So, the equilibrium particle velocity distribution $\psi_0(v)$ is expressed via the pore size distribution determined by Eqs. (11, 12), and vice versa.

More realistic examples for equilibrium particle velocity distribution can be generated using the percolation or effective medium theories [64].

Figure 4a depicts the pore throat size distribution $g(r_p)$ of Berea sandstone as obtained from mercury porosimetry (the data are taken from [2]). The particle velocity distribution $\psi_0(v)$ calculated by Eqs. (11, 12) is shown in Figure 4b. The corresponding normalized particle velocity distribution is $\psi_1(y), y = v/\bar{v}$ with unitary mean, as it is presented in Figure 4c.

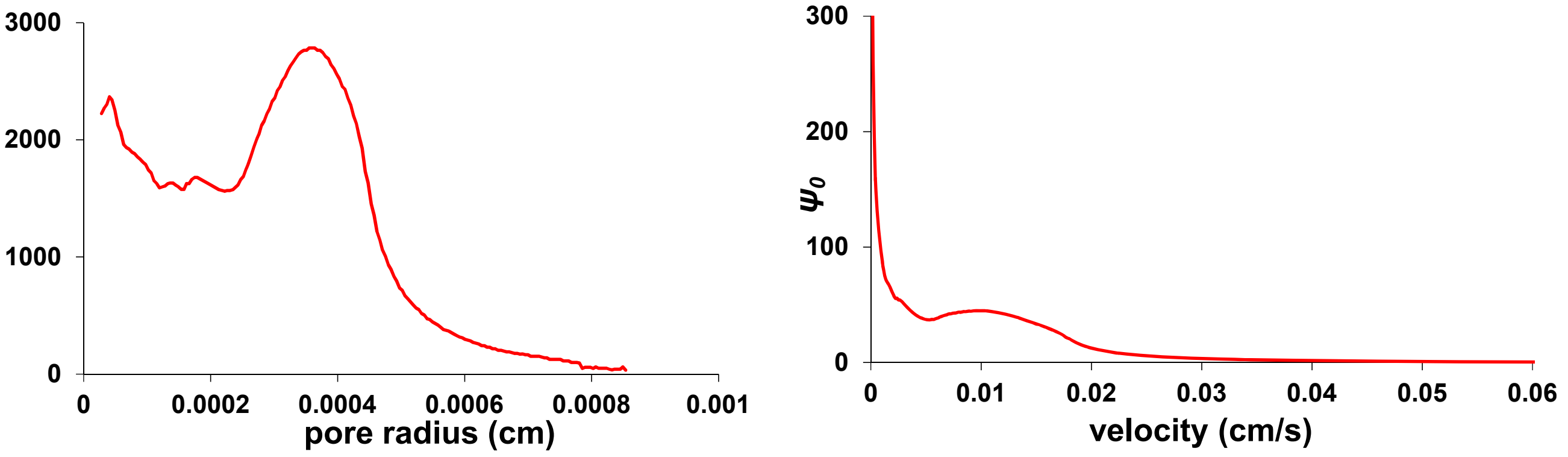

a) b)

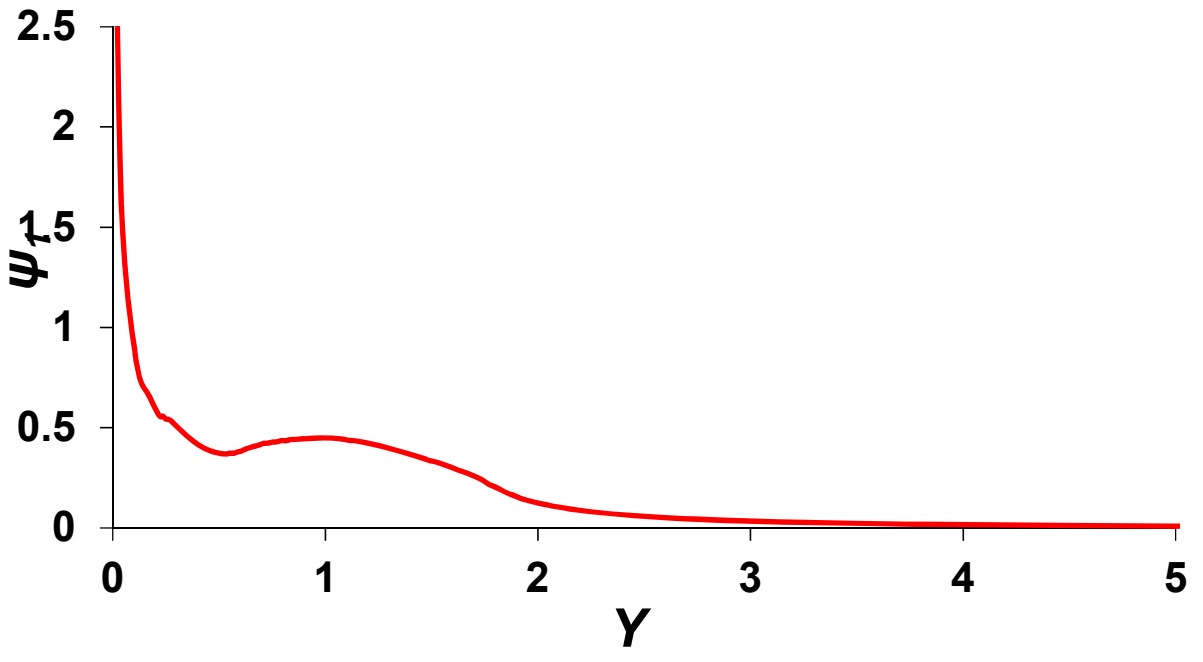

c)

Figure 4. Estimation of velocity distribution for porous media of parallel capillary with mixing chambers: a) pore size distribution; b) equilibrium particle velocity distribution obtained from Pouseuille flow; c) normalised particle velocity distribution.

*Assumption of fixed mixture length* Consider 1D capture-free suspension flow in a model porous medium (Figures 1-3). Assume that the equilibrium distribution was perturbed at some point $(x,t)$. The complete mixing takes place after the particles have reached the next chamber, where the particle size distribution comes back to equilibrium; the delay time is equal to the time of motion from one chamber to another

$$\tau = \frac{l}{\bar{v}} \quad , \tag{13}$$

where $l$ is the inter-chamber distance and $\bar{v}$ is the mean particle velocity at equilibrium

$$\bar{v} = \int_{-\infty}^{\infty} v\psi_0(v)\,dv. \tag{14}$$

Similar speculations can be applied to any 3D porous space. For slow non-inertial flows, the Navier–Stokes equations degenerate into the linear Stokes equations. Consider the phase portrait of particle trajectories in the pore space. From the linearity of the governing equations it follows that an $n$-fold increase in the pressure drop between the inlet and outlet of the flow domain results in the $n$-fold increase in the velocity at each point of the domain, while the particle trajectories remain the same.

The mixing length $l$ in rock with micro-scale heterogeneity depends on the geometry of trajectories. It follows from the statements above that this length is independent of the flow velocity. So, the relaxation time is defined by formula (13) for any pore space topology.

Mathematically, the relaxation of the current distribution towards the equilibrium distribution is expressed by a non-equilibrium equation with linear kinetics and relaxation time $\tau$, which is a simplified version of the BGK (Bhatnagar–Gross–Krook) expression for the collision integral in the Boltzmann equation [14, 43].

Usually, in molecular physics, the relaxation time $\tau$ in a reference volume is constant and can be calculated from the ensemble properties [43]. The above speculations show that the reference constant in a porous medium is the mixing length $l$, which is defined as the correlation length for the stochastically distributed permeability calculated from its variogram – an arbitrary particle velocity distribution becomes equal to $\psi_0(v)$ after the particles have travelled the distance $l$. The relaxation time is not constant; Eq. (13) shows that the relaxation time is reciprocal to the mean flow velocity.

The model assumes that on the micro-scale, the particle capture rate for the size exclusion retention mechanism is proportional to the modulus of the flow velocity, just as at the macro scale in Eq. (4). Unlike first order chemical reactions, obeying the law of acting masses, where the reference time corresponds to the meeting frequency of reacting molecules [39], in porous media it is assumed that the probability $p$ for a particle to be captured during its motion across the unit length is constant [36, 56]. Figure 3 shows the capture vacancies for a particle along its trajectory - thin pore throats that strain the larger particles.

### 2.2. Sink-source term in kinetic equation instead of Cauchy data

Following the assumptions formulated in the previous section, the kinetic equation can be represented in the following form

$$\phi\partial_t f + v\partial_x f = -\lambda|v|f + \bar{v}l^{-1}((\int f\,dv)\psi_0 - f), \tag{15}$$

where $\phi$ is the porosity. Further in the text, integrals in $v$ are taken from minus infinity to infinity. The suspended particles move in the porous space, which explains the appearance of porosity in front of the accumulation term of particle balance (15).

The kinetics term on the right hand side of (15) means that in the absence of capture any particle distribution over velocity, $f(x,t,v)$, is equal to the equilibrium distribution $\psi_0$ after travelling the distance $l$.

The ensemble of particles with equilibrium distribution function $\psi_0$ forms a solution $f(x,t,v)=\psi_0(v)c(x,t)$ for a capture-free flow. The solution corresponds to the set of travelling waves of the overall concentration, where the particles, distributed over velocity, travel with their "own" velocity.

The averaging over velocity (integrating both parts of (15) with respect to $v$ from minus infinity to infinity) and scaling time $t \to t/\phi$ produces the macroscopic transport equation

$$\partial_t c + \partial_x q = -\varepsilon\ , \tag{16}$$

where the capture rate is

$$\varepsilon = \varepsilon(t,x) = \lambda\int|v|f(t,x,v)\,dv\ , \tag{17}$$

and the total particle concentration and flux are given by formulae (7, 8).

The macroscopic transport problem is underdetermined, since Eq. (17) contains three unknown functions: $c=c(t,x)$, $q=q(t,x)$, and $\varepsilon=\varepsilon(t,x)$. In order to make the macroscopic mathematical problem closed, it is necessary to introduce constitutive relations, which are explicit expressions for the mass flux $q$ and particle capture rate $\varepsilon$ through the density $c$ and its derivatives.

The constitutive relations given by Eqs. (8, 17) further in the text are derived from the kinetic equation (15) using the hydrodynamic sink-sources technique, which was previously developed in the kinetic theory of plasma, gases and gas transport in porous media [24-27]. The hydrodynamic sink-source technique requires performing the following derivations:

a) introduce explicit sources in the kinetic equation (15) instead of initial and boundary conditions;
b) obtain expressions for the particle flux $q$ and capture rate $\varepsilon$ in explicit form;
c) derive constitutive relations by eliminating the source function.

A nonzero solution of the kinetic equation (15) corresponds to either an initial condition for the distribution function or a source term in the equation. A particular case of the latter option is an instantaneous source term proportional to the Dirac delta function $\delta(t-0)$. The Cauchy problem subject to an evolution system

$$\frac{\partial u(x,t)}{\partial t} = L_n(u), u(x,0) = u_0(x), \tag{18}$$

can be substituted by the following system

$$\frac{\partial u(x,t)}{\partial t} = L_n(u) + u_0(x)\delta(t-0). \tag{19}$$

The source term in Eq. (19) instantaneously brings the system with any initial condition to state $u_0(x)$ at the moment $t=0+0$.

So, the extended equation (15) with the introduced sink-source term and scaled time $t \to t/\phi$ describes the particle ensemble drift with capture and equilibrium velocity distribution $\psi_0(v)$:

$$\partial_t f + v\partial_x f = -\lambda |v| f + \bar{v}l^{-1}((\int f\,dv)\psi_0 - f) + s(t,x)\psi_0(v). \tag{20}$$

Almost all analytical studies of the Boltzmann equation have been restricted to near equilibrium situations, for which the asymptotic Chapman–Enskog method is applicable and approximate solutions to the equations have been found [14, 15, 43]. The introduction of the sink-source term in the Boltzmann equation (20) instead of the initial non-zero Cauchy data allows exact averaging, which does not require small deviation of the ensample from its equilibrium state, as is shown in the next section.

## 3. Homogenization of the Boltzmann's kinetic equation

In this section, we formulate Boltzmann's equation presented in the previous section, in terms of linear operators in Hilbert space (section 3.1). The averaging is achieved by projection of the solution onto the sub-space of the averaged values, yielding the upscaled transport equation (section 3.2).

### 3.1. Operator form of the Boltzmann´s equation in Hilbert space

In order to solve Eq. (20), it is convenient to perform certain changes of variables and to introduce new mathematical notations. Let us make the following substitution

$$f(x,t,v) = \varphi(x,t,v)\psi_0(v), \tag{21}$$

and treat $\varphi = \varphi(t,x,v)$ as new unknown function.

The substitution of the form of solution (21) into Eq. (20) results in a linear integro-differential equation for the unknown $\varphi = \varphi(t,x,v)$

$$\partial_t \varphi + v\partial_x \varphi = -\lambda |v| \varphi + \bar{v}l^{-1}((\int \varphi\psi_0\,dv) - \varphi) + s(t,x). \tag{22}$$

The solution of Eq. (22) $\varphi=c(x,t)$ corresponds to a set of hyperbolic waves where particles are distributed by equilibrium probability function $\psi_0(v)$. The particles in waves move with "their own velocity" $v$, and the decrease decrement in each wave depends on $v$.

Assume that, with respect to velocity, the function $\varphi = \varphi(t,x,v)$ belongs to the Hilbert space $H$ with scalar product [40]

$$(\phi_1, \phi_2) = \int \psi_0\, \phi_1^* \phi_2\, dv. \tag{23}$$

The kinetic equation for the function $\varphi = \varphi(t,x,v)$, as formulated in terms of the scalar product (23), follows from equation (22):

$$\partial_t \varphi + v\partial_x \varphi + \lambda |v| \varphi - \bar{v}l^{-1}((1,\varphi) - \varphi) = s(t,x). \tag{24}$$

The suspension density, the flux and the sink term can be also expressed in terms of the scalar product (23):

$$c = (1,\varphi) = \int_{-\infty}^{\infty} \psi_0 \varphi\, dv, \tag{25}$$

$$q = (v,\varphi) = \int v\psi_0 \varphi\, dv, \tag{26}$$

$$\varepsilon = \lambda(|v|,\varphi) = \lambda\int |v| \psi_0 \varphi\, dv. \tag{27}$$

This Hilbert space can be regarded as an orthogonal sum

$$H = H_c \oplus H_a, \tag{28}$$

where $H_c$ is a one-dimensional subspace of constants. Let us introduce the following orthogonal projections

$$P_c : H \to H_c,\ P_a : H \to H_a\ , \tag{29}$$

and embeddings

$$J_c : H_c \to H,\ J_a : H_a \to H. \tag{30}$$

In accordance with (25) and (28), any function (vector) $\phi$ can be decomposed as follows:

$$\varphi = c + a,\ c = \int \varphi \psi_0 dv,\ a = \varphi - c\ . \tag{31}$$

where $c$ is an average value of $\varphi$, and $a$ is the difference between $\phi$ and its averaged value, so that the average of $a$ is zero. The projections to the subspaces are

$$c = P_c \varphi,\ \ a = P_a \varphi\ . \tag{32}$$

From the definition (7) it follows that

$$P_c(\varphi) = (1,\varphi),\ P_a(\varphi) = \varphi - (1,\varphi). \tag{33}$$

The embeddings $J_c$ and $J_a$ are representations of the average constant $c$ and the function $a$ as general functions of $v$, respectively.

Applying the Fourier transform to Eq. (22) yields

$$L\,\phi_F = s_F, \tag{34}$$

$$L = i\omega + ikv + \lambda|v| - \tau^{-1}((1,\bullet) - 1)$$
$$g_F(\omega,k) = \int \exp(-i\omega t - ikx)\, g(t,x)\, dt\, dx\ ,$$
$$g(t,x) = (2\pi)^{-2} \int \exp(i\omega t + ikx)\, g_F(\omega,k)\, d\omega\, dk$$

where $g_F(\omega,k)$ correspond to the Fourier transform of an arbitrary function $g = g(t,x)$ along with the notations for inverse Fourier transform [40].

Projecting Eq. (34) into $H_a$ (applying the operator $P_a$ to both sides of Eq. (34)) results in

$$P_a L \varphi_F = P_a s_F, \tag{35}$$

The source function is independent of $v$; hence the function $s(t,x)$ belongs to $H_a$ and so

$$P_a s_F = 0, \tag{36}$$

and the solution $\varphi_F$ belongs to the kernel of operator $P_a L$: $P_a L \varphi_F = 0$.

Using the decomposition (28, 32, 33) of each vector into the sum of its average and the vector with zero average, we decompose the operator $L$ as follows:

$$L = \begin{pmatrix} L_{cc} & L_{ca} \\ L_{ac} & L_{aa} \end{pmatrix}, \tag{37}$$

where the matrix elements are

$$L_{ik} = P_i L J_k \,, \tag{38}$$

$$P_a L = \begin{pmatrix} L_{ac} & L_{aa} \end{pmatrix} : H \to H_a \,. \tag{39}$$

The decomposition of the vector $\varphi_F$ into the sum of $c_F$ and $a_F$ allows rewriting Eqs. (37, 38) as

$$L_{an} n_F + L_{aa} a_F = 0 \,. \tag{40}$$

From Eq. (40) we derive the expression for $a$ as follows

$$a_F = -L_{aa}^{-1} L_{an} c_F \,. \tag{41}$$

The detailed derivations of the operators $L_{ac}$, $L_{aa}$ and $L_{aa}^{-1}$ are presented in Appendix A.

### 3.2 Derivation of constitutive equations

In this section, expressions for the flux $q_F$ and capture rate $\varepsilon_F$ are derived as functions of the overall concentration $c_F$.

Let us first calculate the particle flux in terms of Fourier transforms. Applying the Fourier transform to both sides of Eq. (8) yields

$$\begin{aligned} q_F &= \int v\varphi_F \psi_0 dv = \int v(c_F + a_F)\psi_0 dv = \int v c_F \psi_0 dv + \int v a_F \psi_0 dv = \\ &= c_F \left\langle 1, v \right\rangle + \int v a_F \psi_0 dv \end{aligned} \,. \tag{42}$$

The last term in Eq. (42) is

$$\begin{aligned} &\int v a_F \psi_0 dv = \int \left[ v - \left(1, v\right) + \left(1, v\right) \right] a_F \psi_0 dv = \\ &\int \left[ v - \left(1, v\right) \right] a_F \psi_0 dv + \int \left(1, v\right) a_F \psi_0 dv = \int \left(P_a v\right) a_F \psi_0 dv \end{aligned} \,. \tag{43}$$

Substituting expression (43) into Eq. (42), one obtains the following expression for the particle flux $q_F$

$$\begin{aligned} q_F &= c_F \left\langle 1, v \right\rangle + \int \left(P_a v\right) a_F \psi_0 dv = c_F \left\langle 1, v \right\rangle + \left\langle P_a v, a_F \right\rangle = \\ &= c_F \left\langle 1, v \right\rangle + \left\langle P_a v, -L_{aa}^{-1} L_{ac} c_F \right\rangle = c_F \left\langle 1, v \right\rangle - \left\langle P_a v, L_{aa}^{-1} L_{ac} c_F \right\rangle = \\ &= c_F \left\langle 1, v \right\rangle - \left\langle P_a v, L_{aa}^{-1} \left\{ ik\left(v - \left\langle 1, v \right\rangle\right) + \lambda\left(|v| - \left\langle 1, |v| \right\rangle\right) \right\} c_F \right\rangle = \\ &= c_F \left\{ \left\langle 1, v \right\rangle - ik \left\langle P_a v, L_{aa}^{-1}\left(v - \left\langle 1, v \right\rangle\right) \right\rangle - \lambda \left\langle P_a v, L_{aa}^{-1}\left(|v| - \left\langle 1, |v| \right\rangle\right) \right\rangle \right\} \end{aligned} \,. \tag{44}$$

As follows from (44), the flux $q_F$ can be presented in the form

$$q_F = c_F \left( \bar{v} - ikR_{11} - \lambda R_{12} \right), \tag{45}$$

where

$$\bar{v} = \left\langle 1, v \right\rangle \tag{46}$$

is the average velocity of particles with equilibrium velocity distribution equal to the carrier water velocity, and

$$R_{11} = \left\langle P_a v, L_{aa}^{-1} P_a v \right\rangle, R_{12} = \left\langle P_a v, L_{aa}^{-1} P_a |v| \right\rangle . \tag{47}$$

The expression for the capture rate $\varepsilon_F$ is obtained by the Fourier transform of Eq. (17). The final result is obtained from the expression (45) for flux $q_F$, by substituting $\lambda|v|$ for $v$. Calculations similar to (43, 44) lead to the following expression for the capture rate $\varepsilon_F$

$$\varepsilon_F = \lambda(\overline{|v|} - ikR_{21} - \lambda R_{22})c_F \,, \tag{48}$$

where

$$\overline{|v|} = (|v|,1) = \int |v| \psi_0(v)\,dv \,, \tag{49}$$

is the average of the modulus of the velocity, and

$$R_{21} = (P_a|v|, L_{aa}^{-1} P_a v) \,, \tag{50}$$

$$R_{22} = (P_a|v|, L_{aa}^{-1} P_a |v|) \,. \tag{51}$$

Finally, the expressions for the flux (45) and capture rate (48) are obtained in terms of Fourier transforms; $R_{ij}$ are the transport coefficients.

Now let us calculate the operators $R_{ij}$, $i,j=1,2$

$$R_{ij} = (P_a v_i, L_{aa}^{-1} P_a v_j) \,, \tag{52}$$

where

$$v_1 = v, v_2 = |v| \,. \tag{53}$$

Substituting expression (A-17) for $L_{aa}^{-1}$ into Eq. (52) yields

$$\begin{aligned} R_{ij} &= (P_a v_i, L_{aa}^{-1} P_a v_j) = \left(P_a v_i, \Lambda^{-1} P_a v_j - \Lambda^{-1} \langle 1, \Lambda^{-1} \rangle^{-1} \langle \Lambda^{-1}, P_a v_j \rangle \right) \\ L_{aa}^{-1} &= \Lambda^{-1} - \Lambda^{-1} \langle 1, \Lambda^{-1} \rangle^{-1} \langle \Lambda^{-1}, \bullet \rangle \\ L_{aa}^{-1} P_a v_j &= \Lambda^{-1} P_a v_j - \Lambda^{-1} \langle 1, \Lambda^{-1} \rangle^{-1} \langle \Lambda^{-1}, P_a v_j \rangle \end{aligned} \,. \tag{54}$$

Explicit expressions for the transport coefficients $R_{ij}$ are obtained by substituting expression (A-8) for $\Lambda$ into Eq. (54).

The application of the inverse Fourier transform to expression (45) for the flux results in the expression for the flux in real space

$$\begin{aligned} q_F &= (A_1 - ikR_{11} - \lambda R_{12})c_F = A_1 c_F - R_{11}(ikc_F) - \lambda R_{12} c_F \\ q &= \overline{v} n - K_{11} * \partial_x c - \lambda K_{12} * c \end{aligned} \,, \tag{55}$$

where $K_{11}$ and $K_{12}$ are the inverse Fourier transforms of $R_{11}$ and $R_{12}$, respectively, and a star denotes the convolution in space-time.

Expressions (55) shows that in real space-time the constitutive relations are non-local

$$I = (\overline{v} - \lambda K_{12} *) n - K_{11} * \partial_x n \,, \tag{56}$$

which implies that $K_{11}$ and $K_{12}$ are integro-differential operators.

The expression for the capture rate in real space-time is obtained by applying the inverse Fourier transform to expression (48):

$$\varepsilon = \lambda(\overline{|v|} - \lambda K_{22}{*})n - \lambda K_{21} * \partial_x n . \tag{57}$$

In the case of long waves and large times, where $\omega, k \to 0$, the expressions for $R_{ij}$ (54) become scalars, convolution degenerates into multiplication and $R_{ij}=K_{ij}$. The expressions (B-8) for $R_{ij}$ are derived in Appendix B. The $R_{ij}$-values are expressed via five constants $B_{00}$, $B_1$, $B_2$ , $B_{12}$ and $B_{11}$.

The transport coefficient matrix is symmetric, so $R_{12}=R_{21}$.

Finally, the substitution of expressions (B-14-17) for $R_{ij}$ into (56, 57) leads to the constitutive relations for the flux

$$I = (\overline{v} - \lambda(B_{12} - B_{00}^{-1} B_1 B_2))c - (B_{11} - B_{00}^{-1} B_1^{\,2})\partial_x c = (\overline{v} - \lambda R_{12})c - R_{11}\partial_x c , \tag{58}$$

and the capture rate

$$\varepsilon = \lambda(\overline{|v|} - \lambda(B_{11} - B_{00}^{-1} B_2^{\,2}))c - \lambda(B_{12} - B_{00}^{-1} B_1 B_2)\partial_x c = \lambda(\overline{|v|} - \lambda R_{22})c - \lambda R_{21}\partial_x c . \tag{59}$$

## 4. Large-scale phenomenological model for deep bed filtration

The constitutive relations obtained in previous section allow closing the governing system and formulating the averaged equation for suspension transport in porous media (section 4.1). The difference between the obtained averaged equation and the classical deep bed filtration model is discussed in this section. The properties of three coefficients of the averaged equation are analyzed (section 4.2).

### 4.1. Governing equations on macro scale and modification of the deep bed filtration model

Expressions (58, 59) allow the formulation of the phenomenological model for deep bed filtration. The model consists of the particle balance equation (16) with explicit expressions for the particle flux (58) and capture rate (59). Substituting (58, 59) into Eq. (18) results in a linear advective-diffusion equation with a sink term

$$\partial_t c + \left[\overline{v} - 2\lambda(B_{12} - B_{00}^{-1} B_1 B_2)\right]\partial_x c = (B_{11} - B_{00}^{-1} B_1^{\,2})\partial_{xx} c - \lambda(\overline{|v|} - \lambda(B_{11} - B_{00}^{-1} B_2^{\,2}))c \ . \tag{60}$$

The transport equation (60) can also be expressed in terms of $R_{ij}$. It follows from expressions (56, 57) that

$$\begin{aligned} & q = \left(\overline{v} - \lambda R_{12}\right)c - \lambda R_{11}\partial_x c, \quad \varepsilon = \lambda\left(\overline{|v|} - \lambda R_{22}\right)c - \lambda R_{21}\partial_x c \\ & \partial_t c + (\overline{v} - 2\lambda R_{12})\partial_x c = R_{11}\partial_{xx} c - \left[\lambda\left(\overline{|v|} - \lambda R_{22}\right)\right]c \end{aligned} . \tag{61}$$

The model contains the advective velocity of the particle flux, the particle diffusion and capture intensity. Let us analyze the transport coefficients.

The velocity $\overline{v}$ is the mean velocity, i.e. the velocity of the carrier water. The constants $R_{11}$, $R_{12}$ and $R_{22}$ depend on the equilibrium particle velocity distribution $\psi_0(v)$, reference mixture time $l/\overline{v}$, and micro-scale filtration coefficient $\lambda$.

The proof in Appendix C shows that all transport coefficients and also average velocity modulus are proportional to the mean particle velocity. The ratios of the coefficients to the mean velocity depend on the normalized particle velocity distribution $\psi_1(y)$, $y = v/\overline{v}$. So, the three dimensionless numbers (64) also depend on the normalized velocity distribution $\psi_1(y)$ and are independent of the mean particle velocity.

Introduce the dimensionless coordinate and time

$$T = \frac{t\overline{v}}{\phi L}, X = \frac{x}{L}, \tag{62}$$

and substitute them into the flow equation (61):

$$\frac{\partial c}{\partial T}+(1-\frac{2\lambda R_{12}}{\overline{v}})\frac{\partial c}{\partial X}=\frac{R_{11}}{\overline{v}L}\frac{\partial^2 c}{\partial X^2}-\frac{L}{\overline{v}}\left[\lambda(\overline{|v|}-\lambda R_{22})\right]c\ . \tag{63}$$

Eq. (61) contains three dimensionless macroscale parameters of delay, diffusion and capture:

$$\theta=\frac{2\lambda R_{12}}{\overline{v}},\frac{1}{Pe}=\frac{R_{11}}{\overline{v}L},\Omega=\lambda L(\frac{\overline{|v|}}{\overline{v}}-\frac{\lambda R_{22}}{\overline{v}})\ . \tag{64}$$

They depend on two dimensionless micro-scale parameters of mixing length ratio *l/L*, filtration-mixing number *λl*, and the equilibrium velocity distribution $\psi_0(v)$. Two-parametric velocity distributions are determined by the average velocity $\overline{v}$ and the coefficient of variation $C_v$.

### 4.2. Properties of upscaled model coefficients

***Dimensionless diffusion 1/Pe.*** The expression for the Schmidt number (the inverse of the Péclet number) follows from expression (C-2)

$$\frac{1}{Pe}=\frac{1}{\lambda L}\left[\int\frac{y^2}{|y|+1/\lambda l}\psi_1(y)\,dy-\frac{\left(\int\frac{y}{|y|+1/\lambda l}\psi_1(y)\,dy\right)^2}{\int\left(|y|+1/\lambda l\right)^{-1}\psi_1(y)\,dy}\right]. \tag{65}$$

The Péclet number depends on the normalized equilibrium particle velocity distribution $\psi_1(v)$, the dimensionless filtration coefficient *λL,* and the dimensionless mixture length *l/L*. It is independent of the mean flow velocity.

The expression in square brackets in Eq. (65) depends on the product of the dimensionless parameters *λL* and *l/L*, i.e. *λl*, which is further called the filtration-mixing number. The ratio of the dimensionless filtration coefficient *λL* to the Péclet number depends on *λl* only.

Let us prove that the diffusion coefficient is always positive.

*Lemma 1.* Either of the following inequalities holds

$$R_{11}>0 \text{ or } B_{11}>B_{00}^{-1}B_1^{\ 2}\ . \tag{66}$$

*Proof.* Eq. (B-14) shows the equivalence of the two statements. Determine the scalar product in Hilbert space of continuous differentiable functions as

$$\langle x,z\rangle=\int\frac{xz\psi_0(v)\,dv}{\lambda|v|+\overline{v}/l}\ . \tag{67}$$

From the Cauchy–Bunyakovsky-Schwarz inequality [40] it follows that

$$\langle x,x\rangle\langle z,z\rangle\ge\langle x,z\rangle^2,\ \langle y,y\rangle\langle 1,1\rangle\ge\langle 1,y\rangle^2\ . \tag{68}$$

For the scalar product (67), expressions (B-9-13) give

$$B_{11}=\langle v,v\rangle,B_{00}=\langle 1,1\rangle,B_{12}=\langle 1,v\rangle\ . \tag{69}$$

Therefore,

$$B_{11}>B_{00}^{-1}B_1^{\ 2}, \tag{70}$$

which proves the lemma.

The dependence of *1/Pe* on the filtration coefficient $\lambda L$ expresses the effect of particle capture on diffusion. Figure 5 shows the plot of *1/Pe* versus $\lambda L$ for normalized normal velocity distribution $\psi_1(y)$ and three different values for coefficient of variation $C_v$= 0.01, 0.1 and 1.0 (Figures 5a, b, c, respectively) displayed in log-log coordinates. In the expression (65) for upscaled inverse Peclet's number, $\lambda L$ and $l/L$ are independent dimensionless groups, and $\lambda l$ is their product. The typical values of the filtration coefficient, depending on the particle size, mineralogy of the particles and matrix, their zeta-potential, water pH and salinity, and the flow velocity, vary between 0.01 $m^{-1}$ (for bacteria in high-permeability aquifers) and 100 $m^{-1}$ (for seawater injection in low-permeability oilfield) [36]. The length $L$ varies from 0.01 m for short reservoir cores to 1 m for outcrop specimen and up to 100 for natural reservoirs. The mixing length $l$ varies from $10^{-4}$ m for highly-sorted high-permeability sandstones up to 0.1 m for low-permeable highly heterogeneous cores and up to 10 m for heterogeneous reservoirs [2, 4]. So, the dimensionless filtration coefficient $\lambda L$ in Figure 5 varies from $10^{-3}$ to $10^{4}$, and typical values for $l/L$ are taken as 0.01, 0.1 and 1.0.

As is expected, the higher the velocity variation coefficient the higher the diffusion. Each change of $C_v$ from 0.01 to 0.1 and from 0.1 to 0.01 results in a 100-fold increase in *1/Pe*.

Since particle capture is proportional to the velocity magnitude, capture decreases the velocity standard deviation and, consequently, decreases particle diffusion. Therefore, the larger the filtration coefficient the lower the diffusion. The maximum value for dimensionless diffusion is obtained from formula (65) for $\lambda$=0; it is equal to $\frac{l}{L}\left[\int y^2\psi_1(y)\,dy-1\right]$. Hence, the parameter 1/*Pe* decreases from its maximum value to zero for $\lambda L$ increasing from zero to infinity. Figure 5 shows that 1/*Pe* almost vanishes for $\lambda L$ exceeding 100. So, particle capture strongly affects diffusion in the large-scale model (63), while the classical deep bed filtration model given by Eqs. (1-4) assumes that the diffusion coefficient is independent of the filtration coefficient.

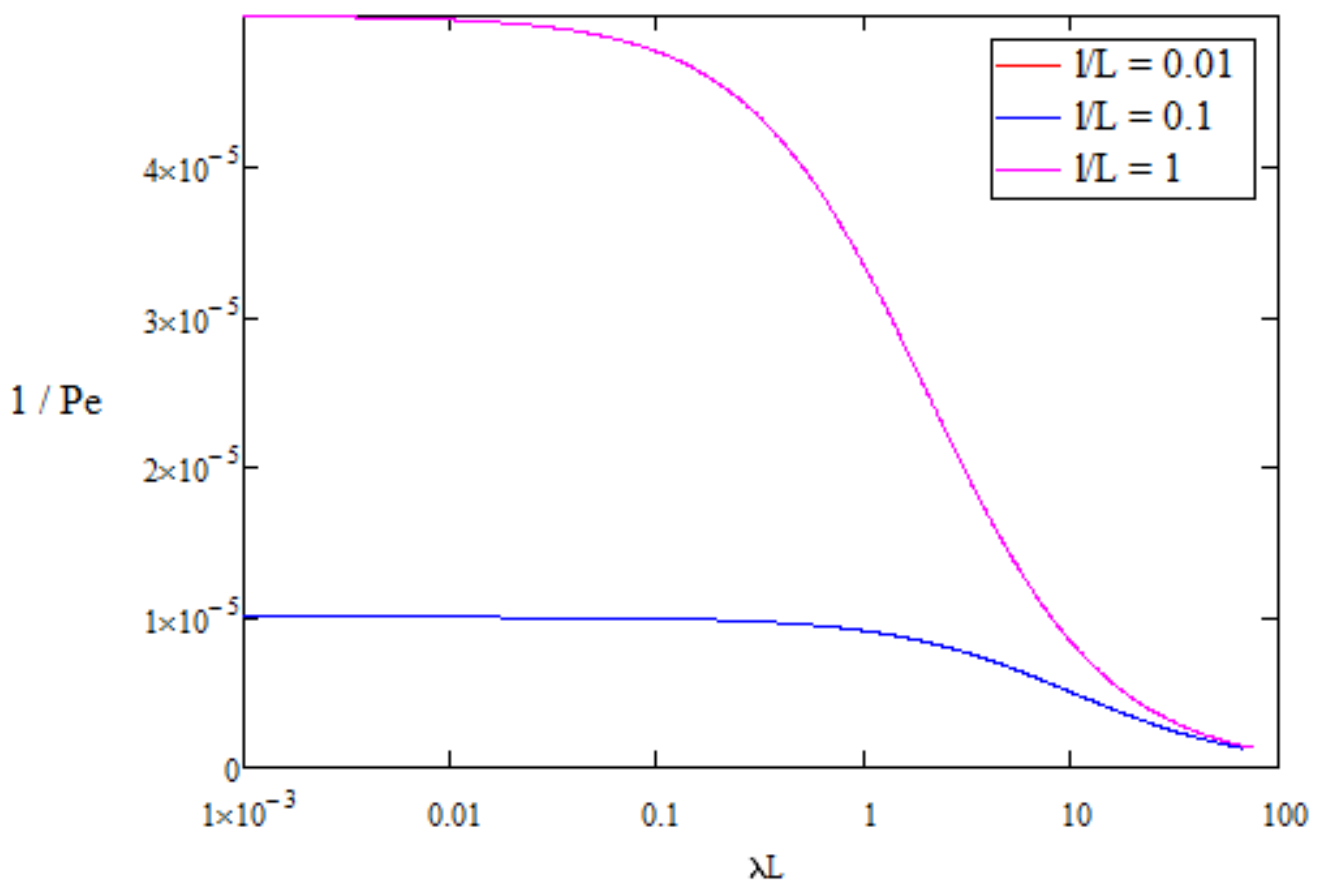

a)

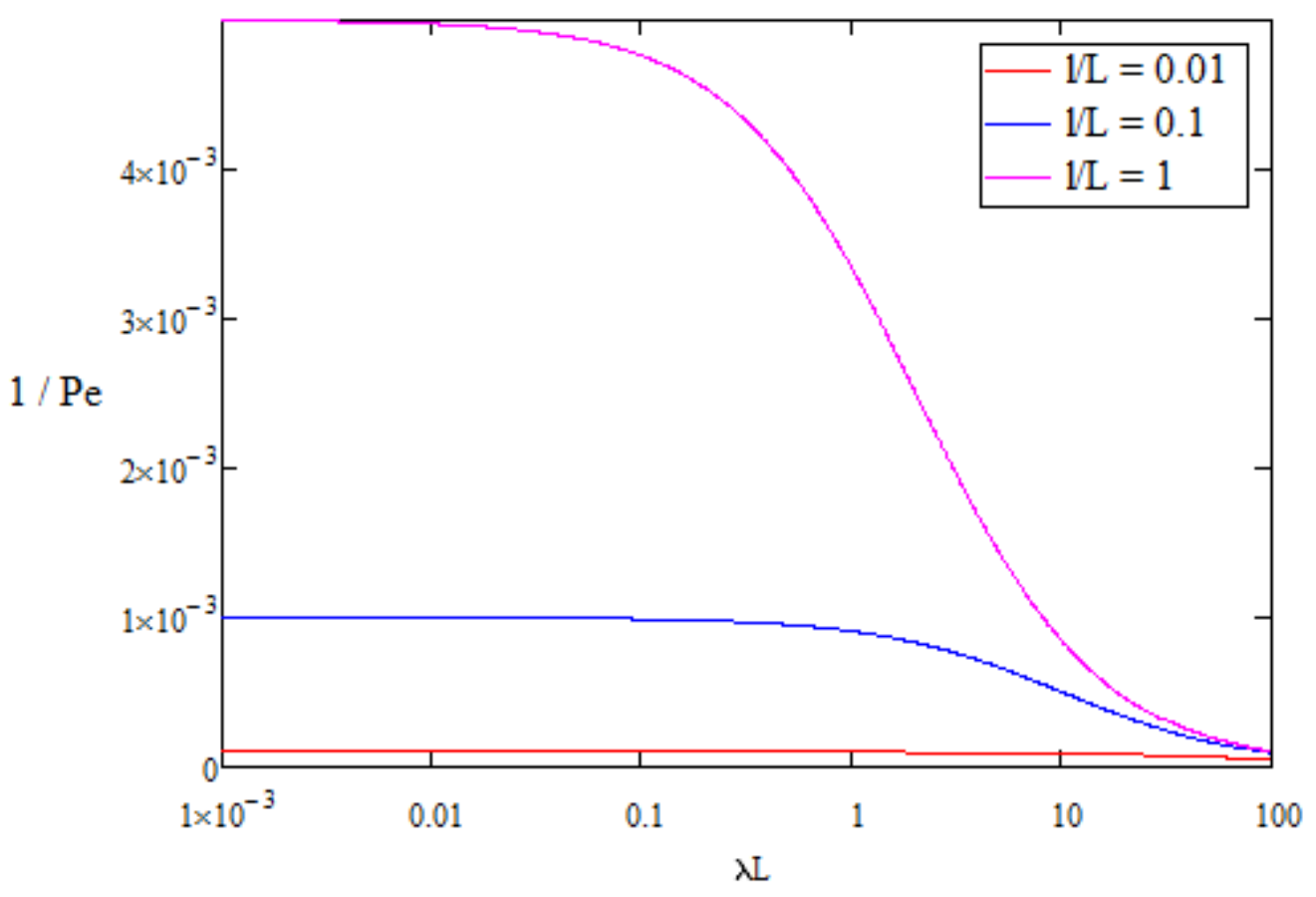

b)

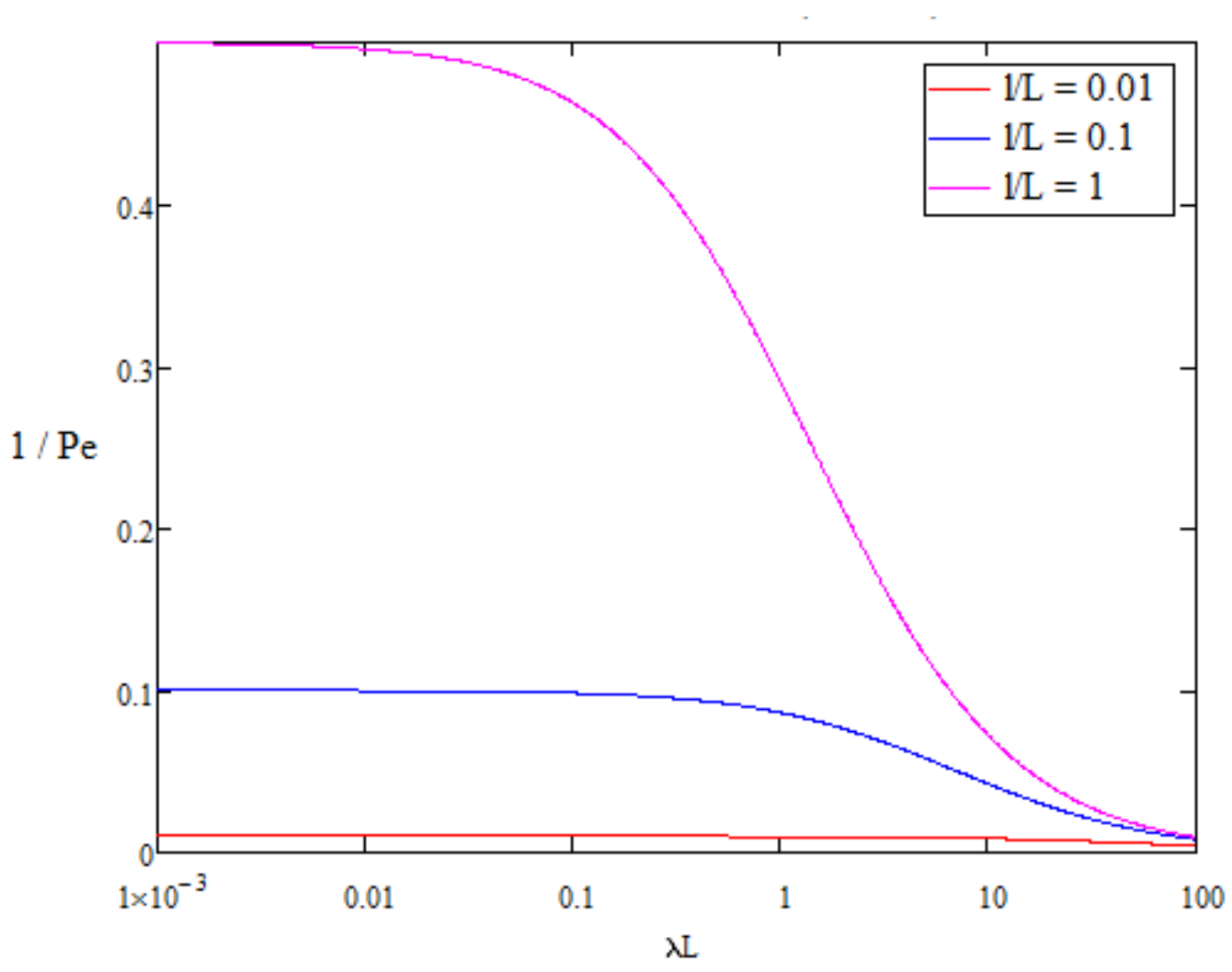

c)

Figure 5. Effect of particle capture on dispersion – filtration-coefficient dependence of reciprocal Peclet number: a) coefficient of variation $C_v$=0.01; b) $C_v$=0.1; c) $C_v$=1.0.

Figure 6 shows the ratio between the upscaled reciprocal to Peclet number *1/Pe* and dimensionless micro-scale mixing length *l/L*, which is equal to upscaled and micro-scale dispersivity coefficients; here *1/Pe* is calculated by Eq. (65) for a normalized normal velocity distribution as per

$$\frac{L}{Pe\times l}=\frac{1}{\lambda l}\left[\int\frac{y^2}{|y|+{}^1\!/_{\lambda l}}\psi_1(y)\,dy-\frac{\left(\int\frac{y}{|y|+{}^1\!/_{\lambda l}}\psi_1(y)\,dy\right)^2}{\int\left(|y|+{}^1\!/_{\lambda l}\right)^{-1}\psi_1(y)\,dy}\right]. \qquad (71)$$

The upscaled reciprocal to the Peclet number depends on the boundary problem size *L*, whereas the ratio (71) is independent of *L*. Therefore, the independent variable during calculations in Figure 6 is the filtration-mixing number $\lambda l$, which is the ratio between the mixing length and the capture-free run. The capture rate in Eq. (15) is proportional to module of velocity, so the capture eliminates mostly "fast" particles, resulting in decrease in the coefficient of variation of velocity and dispersivity; this explains the curve decline in Fig. 6. Assuming typical values for the filtration coefficient, which reflects the particle capture rate at the micro-scale, and the mixing length for dispersed micro-scale heterogeneous system, the parameter $\lambda l$ varies from $10^{-5}$ (small particles in homogeneous cores) to 10 (large particles in heterogeneous specimen). The ratio varies from the coefficient of variation $C_v$ squared at $\lambda l$=0 to zero where $\lambda l$ tends to infinity.

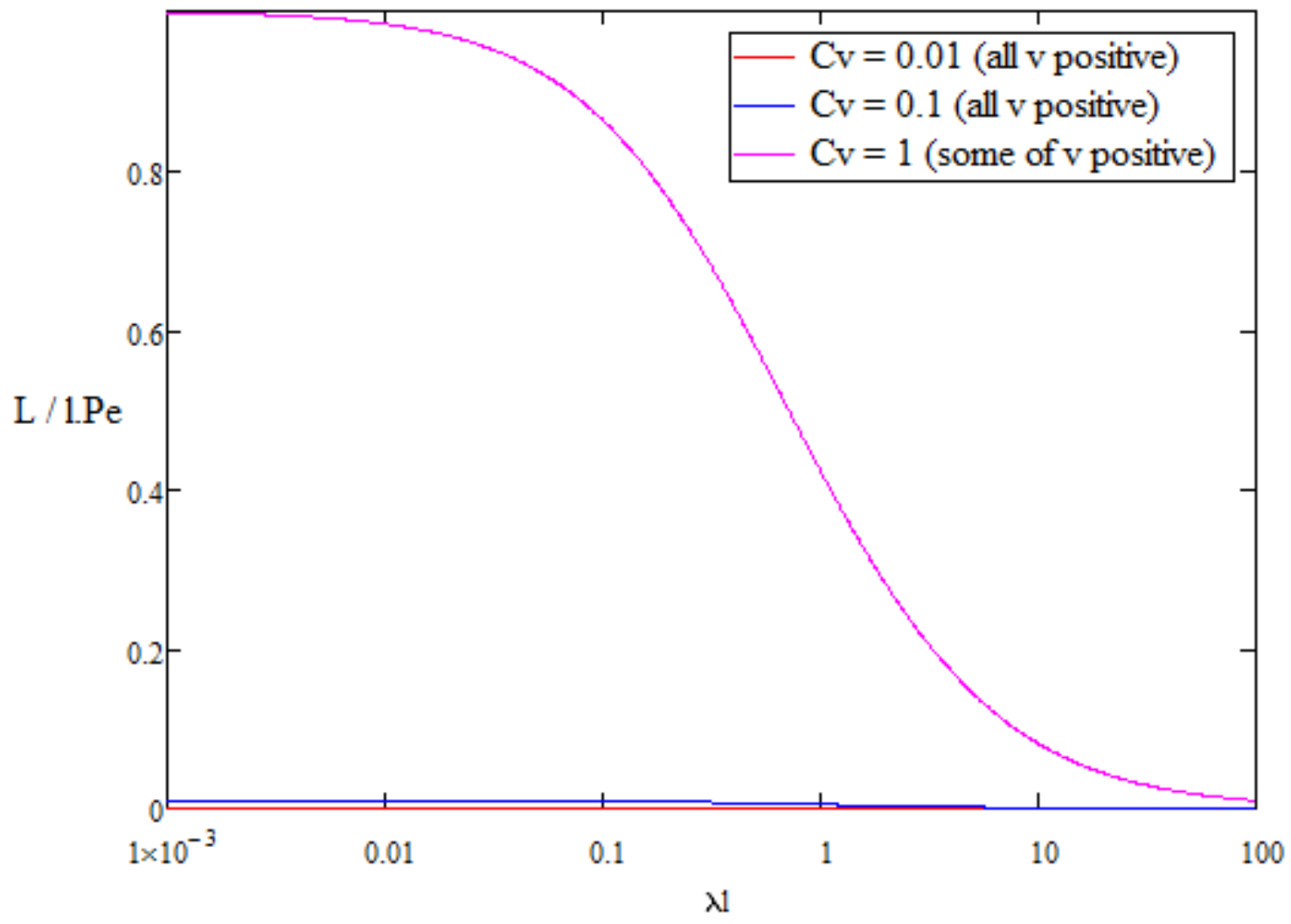

Figure 6. Effect of particle capture on the ratio between the dispersivity coefficient and mixing length for coefficients of variation $C_v$=0.01, $C_v$=0.1, and $C_v$=1.0.

The plots for *1/Pe* versus $\lambda L$ and *L/(l*Pe)* versus $\lambda l$ for log-normal velocity distributions with the same variation coefficients are almost the same as those shown in Figures 5 and 6. However, the form of the plot of *1/Pe* versus $\lambda L$ for Berea sandstone is different. The dimensionless diffusion versus dimensionless capture in Figure 7 is calculated for the porous medium properties presented in Figure 4.

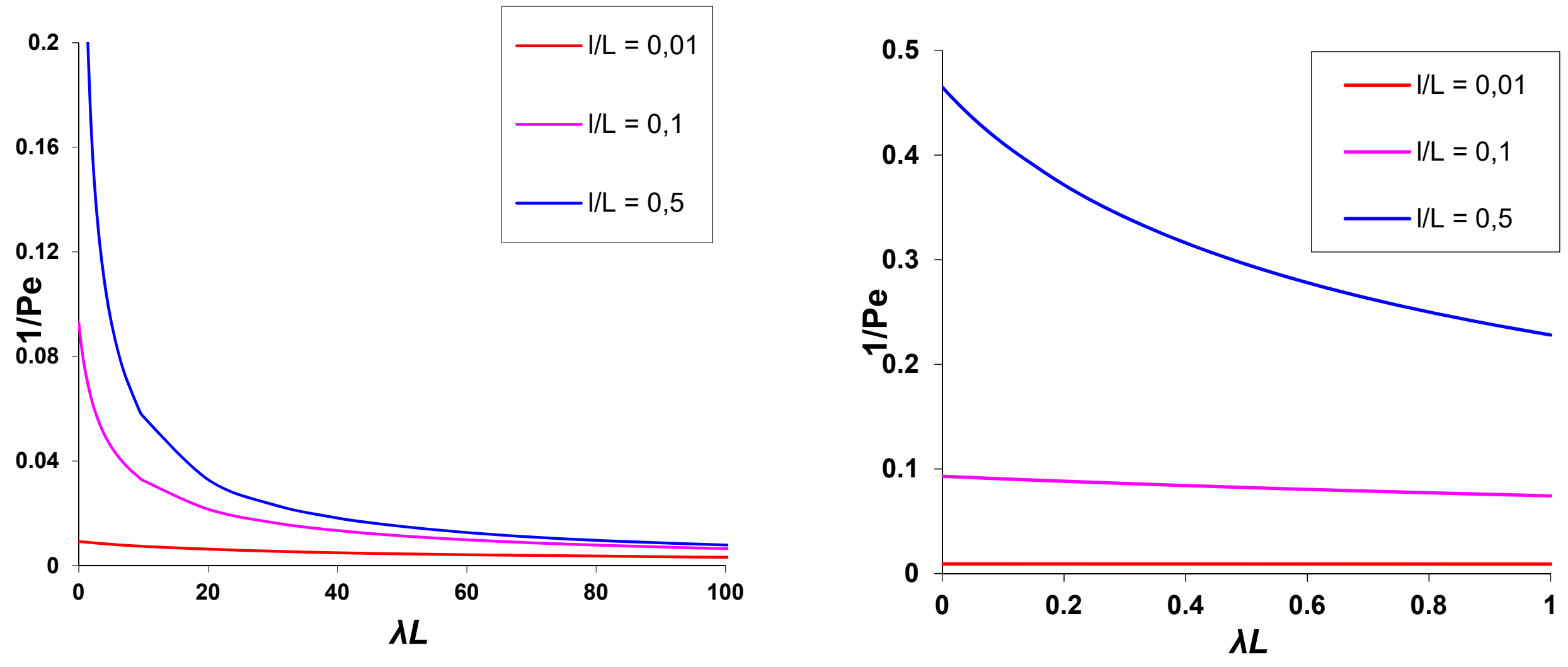

a) b)

Figure 7. Dependence of dimensionless diffusion (inverse to Peclet number) of dimensionless filtration coefficient $\lambda L$ for different mixing lengths *l/L*: a) interval of filtration coefficients for core and field scales; b) zoom for core scale.

***Delay parameter*** $\boldsymbol{\theta}$ The expression for the delay number $\theta$ is obtained by substituting Eq. (C-2) into expression (64)

$$\theta = 2\int \frac{y|y|}{|y|+1/\lambda l}\psi_1(y)\,dy - \frac{2\left(\int \frac{y}{|y|+1/\lambda l}\psi_1(y)\,dy\right)\left(\int \frac{|y|}{|y|+1/\lambda l}\psi_1(y)\,dy\right)}{\int \left(|y|+1/\lambda l\right)^{-1}\psi_1(y)\,dy}. \tag{72}$$

The delay number depends on the product of the dimensionless filtration coefficient $\lambda L$ and dimensionless mixture length *l/L*, which is the filtration-mixing number $\lambda l$, and on the normalized equilibrium velocity distribution $\psi_1(v)$. It is independent of the mean flow velocity.

The capture rate on the micro-scale is proportional to the magnitude of the particle velocity, as it is in Eq. (15). Hence the capture excludes preferentially fast particles from the flux, which means that capture with the rate expression in Eq. (15) slows down the flux. Therefore, it is expected that the particle capture and diffusion always introduce delay into the carrier water flux. Thus the value $\theta$ must always be positive.

*Lemma 2.* The following inequalities hold

$$\begin{aligned} &R_{12} = B_{12} - B_{00}^{-1}B_1B_2 > 0 \\ &\theta(\lambda l) > 0 \end{aligned}. \tag{73}$$

The capture rate on the micro-scale is proportional to the magnitude of the particle velocity, as it is in Eq. (15). Hence the capture excludes preferentially fast particles from the flux, which means that capture with the rate expression in Eq. (15) slows down the flux. Therefore, it is expected that the particle capture and diffusion always introduce delay into the carrier water flux. Thus the value $\theta$ must always be positive. Here we do not prove this statement but illustrate it by numerical calculations.

The equivalence of two inequalities (73) follows from Eqs. (72) and (B-15). The delay parameter $\theta$ given by the expression (72) is a monotonically increasing function of $\lambda l$, $\theta^{/}(\lambda l)>0$. Because the capture decelerates the particle flux, the larger are the filtration coefficient and the mixing length, the larger is the delay.

Tending filtration-mixing number $\lambda l$ to zero in Eq. (72) shows that each term becomes asymptotically proportional to $\lambda l$ and tends to zero as

$$2\lambda l(\int y|y|\psi_1(y)dy - \int y\psi_1(y)dy \times \int |y|\psi_1(y)dy) . \tag{74}$$

So, the delay parameter vanishes with negligible capture or diffusion. Therefore, in the diffusion-free and capture-free cases, the mean particle velocity is equal to the carrier water velocity.

If $\lambda l$ tends to infinity for high capture rates and diffusion, the term $1/\lambda l$ can be neglected as compared with either $y$ or $|y|$; the first term in (72) tends to two. The delay parameter $\theta$ tends to

$$2\left(1-\frac{\int y|y|^{-1}\psi_1(y)dy}{\int |y|^{-1}\psi_1(y)dy}\right). \tag{75}$$

The delay parameter $\theta$ tends to two for the case where $\psi_1(0)>0$. Otherwise $\theta$ tends to a number lower than two.

Figure 8 shows an increase in the delay parameter as the non-dimensional parameter $\lambda l$ increases. The higher is the coefficient of variation $C_v$, the larger is the fraction of large velocities, the higher is the capture and the higher is the delay. For $\lambda l$=1 and $C_v$=1, the delay parameter reaches the value 0.7, and the mean particle velocity becomes $0.3\overline{v}$ . So, the particle capture significantly slows down the particle flux as compared with the water velocity.

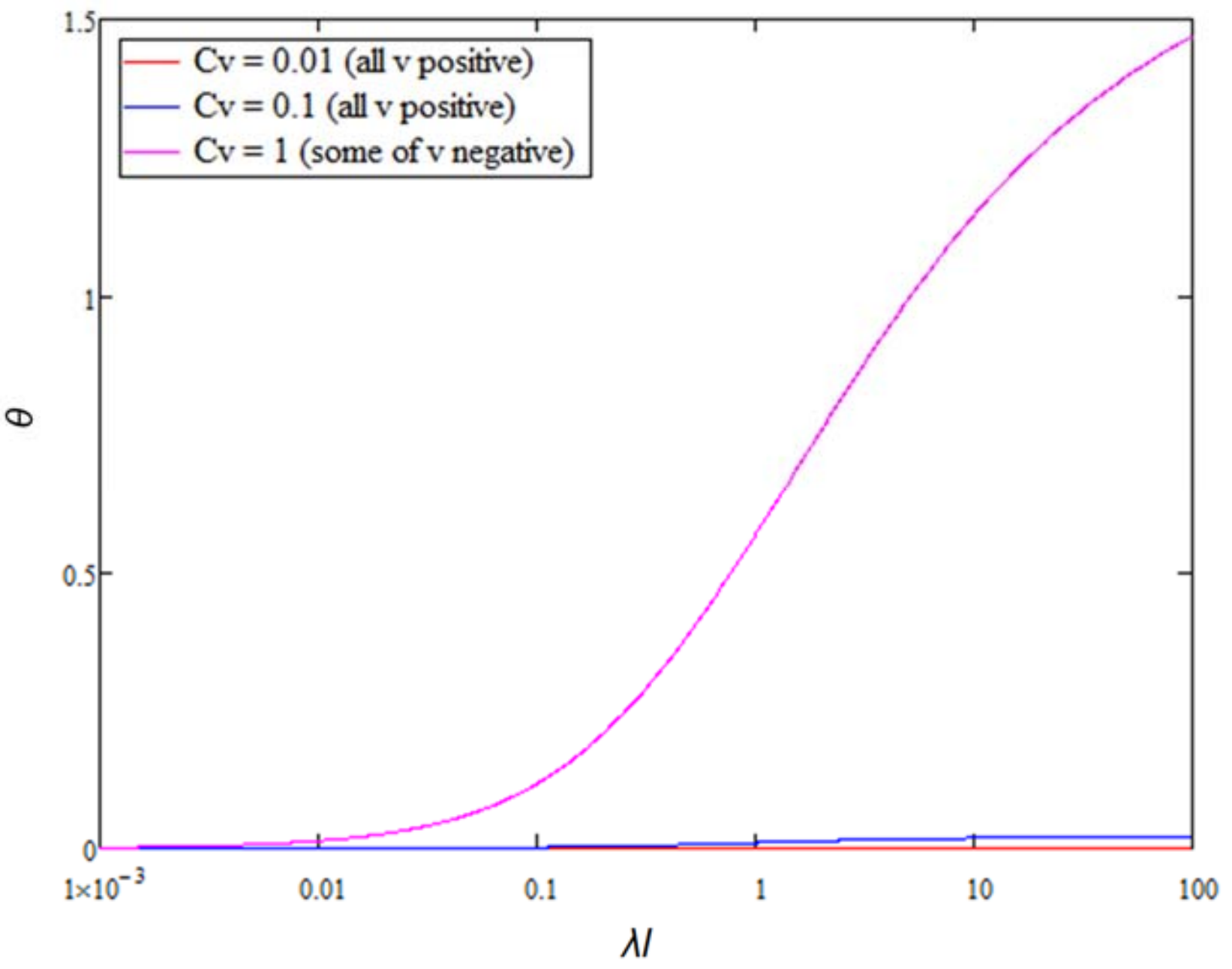

Figure 8. Effect of filtration-mixture number $\lambda l$ on delay $\theta$ at different variance coefficients.

For log-normal velocity distribution, the plot of delay versus $\lambda l$ is almost the same as that shown in Figure 8. Figure 9 shows the delay plot for the Berea core, which properties are presented in Figure 4. The form of the graph is different from that for normal and log-normal velocity distributions.

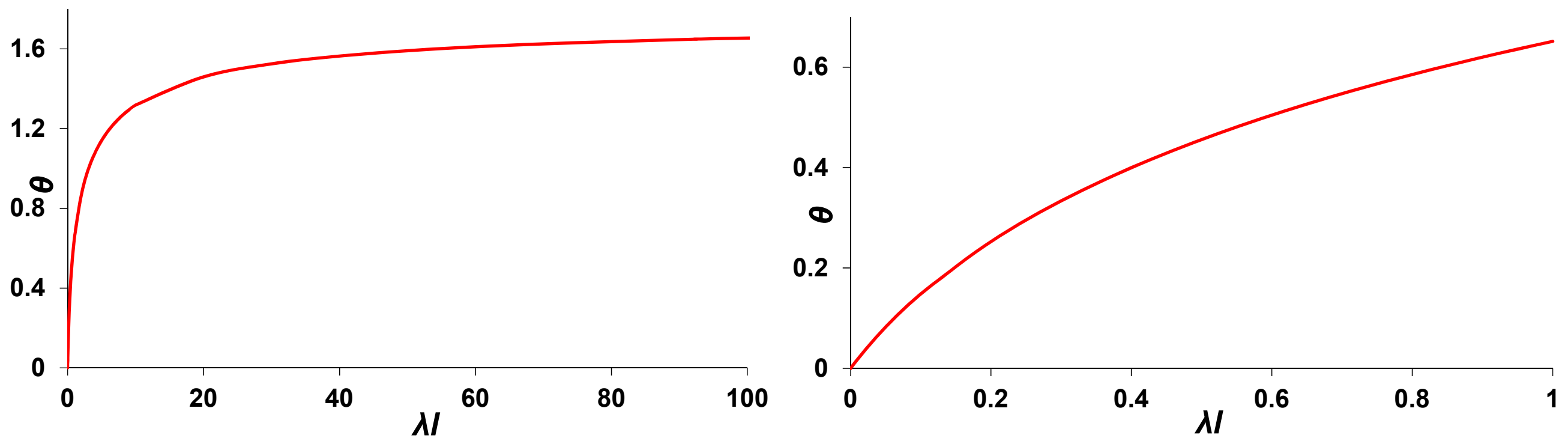

a) b)

Figure 9. The higher is the filtration-mixing number $\lambda l$ the larger is the delay of particles $\theta$ if compared with water: a) dependence of delay parameter of filtration-mixing number $\lambda l$; b) zoom for core scale.

***Dimensionless capture Ω.*** Substituting expression (C-2) into the expression for the dimensionless capture (64) leads to

$$\Omega = \frac{L}{\overline{v}}\left[\lambda(\overline{|v|} - \lambda R_{22})\right] = \lambda L\left[\int |y|\psi_1(y)\,dy - \int \frac{y^2}{|y| + \frac{1}{\lambda l}}\psi_1(y)\,dy + \frac{\left(\int \frac{|y|}{|y| + \frac{1}{\lambda l}}\psi_1(y)\,dy\right)^2}{\int \left(|y| + \frac{1}{\lambda l}\right)^{-1}\psi_1(y)\,dy}\right]. \tag{76}$$

So, the dimensionless capture coefficient depends on the dimensionless filtration coefficient $\lambda L$, its product with dimensionless mixture length $l/L$, and the normalized equilibrium velocity distribution $\psi_1(y)$. It is independent of the mean flow velocity.

*Lemma 3.* Either of the following inequalities holds

$$R_{22}>0, \text{ or } B_{11} > B_{00}^{-1}B_2^2 . \tag{77}$$

*Proof.* Define the scalar product by Eq. (67). From the Cauchy–Bunyakovsky-Schwarz inequality (68) it follows that

$$\langle |v|,|v|\rangle\langle 1,1\rangle \geq \langle 1,|v|\rangle^2 . \tag{78}$$

With the scalar product (67), expressions (B-13, 9, 11) become

$$B_{11} = \langle |v|,|v|\rangle, B_{00} = \langle 1,1\rangle, B_2 = \langle 1,|v|\rangle . \tag{79}$$

Therefore,

$$B_{11}B_{00} > B_2^{\,2} , \tag{80}$$

which proves the lemma.

According to the classical deep bed filtration theory, particle capture is proportional to the absolute value of the mean flow velocity, which is reflected in Eq. (4). Eq. (15) shows that the assumption that the capture rate is proportional to module of velocity holds in the micro-scale Boltzmann model. For the case of positive velocities, the averaging effectively results in the subtraction of the term $\lambda R_{22}$ from the mean flow velocity in the scalar term of the capture, as Eqs. (60) and (61) show. Lemma 3 shows that the subtracted term is always positive, and hence the classical deep bed filtration theory overestimates the scalar term of the capture intensity.

With the diffusion (mixture length *l*) tending to zero, the second and third terms in Eq. (76) also tend to zero. Then the capture number $\Omega$ tends to $\lambda L$ for the case of positive velocities only. Otherwise, the limit is positive and lower than $\lambda L$.

Since there is no particle release in the model (15), the capture rate must always be positive. Let us prove it.

*Lemma 4.* The averaged filtration coefficient is always positive

$$\Omega = \frac{L}{\bar{v}}\left[\lambda(|v| - \lambda R_{22})\right] > 0 . \tag{81}$$

*Proof.* Let us calculate the sum of the first two terms in brackets in Eq. (76)

$$\int |y| \psi_1(y)\,dy - \int \frac{y^2}{|y| + 1/\lambda l}\psi_1(y)\,dy = \int \frac{y^2 + |y|/\lambda l - y^2}{|y| + 1/\lambda l}\psi_1(y)\,dy = \\ = \int \frac{|y|/\lambda l}{|y| + 1/\lambda l}\psi_1(y)\,dy \tag{82}$$

Then the total expression in brackets in Eq. (76) becomes

$$\int |y| \psi_1(y)\,dy - \int \frac{y^2}{|y| + 1/\lambda l}\psi_1(y)\,dy + \frac{\left(\int \frac{|y|}{|y| + 1/\lambda l}\psi_1(y)\,dy\right)^2}{\int \left(|y| + 1/\lambda l\right)^{-1}\psi_1(y)\,dy} = \\ \frac{1}{\lambda l}\int \frac{|y|}{|y| + 1/\lambda l}\psi_1(y)\,dy + \frac{\left(\int \frac{|y|}{|y| + 1/\lambda l}\psi_1(y)\,dy\right)^2}{\int \left(|y| + 1/\lambda l\right)^{-1}\psi_1(y)\,dy} > 0 \tag{83}$$

which is the sum of two positive terms. This proves the lemma.

The dimensionless capture $\Omega$ in Eq. (76) is "almost" proportional to the dimensionless filtration coefficient $\lambda L$, so the larger is the micro-scale filtration coefficient $\lambda l$, the larger is the micro-scale filtration coefficient $\Omega$. The dimensionless capture vanishes as the filtration coefficient tends to zero.

The ratio between the upscaled and pore-scale capture terms $\Omega/\lambda L$ versus $\lambda l$ is shown in Figure 10 for normal velocity-distribution and in Figure 11 for the Berea core. The capture ratio $\Omega/\lambda L$ is a monotonically decreasing function of $\lambda l$. With $\lambda l$ varying from zero to infinity, $\Omega/\lambda L$ varies in the interval

$$\left\{\int |y| \psi_1(y)\,dy,\ \left[\int |y|^{-1}\psi_1(y)\,dy\right]^{-1}\right\} . \tag{84}$$

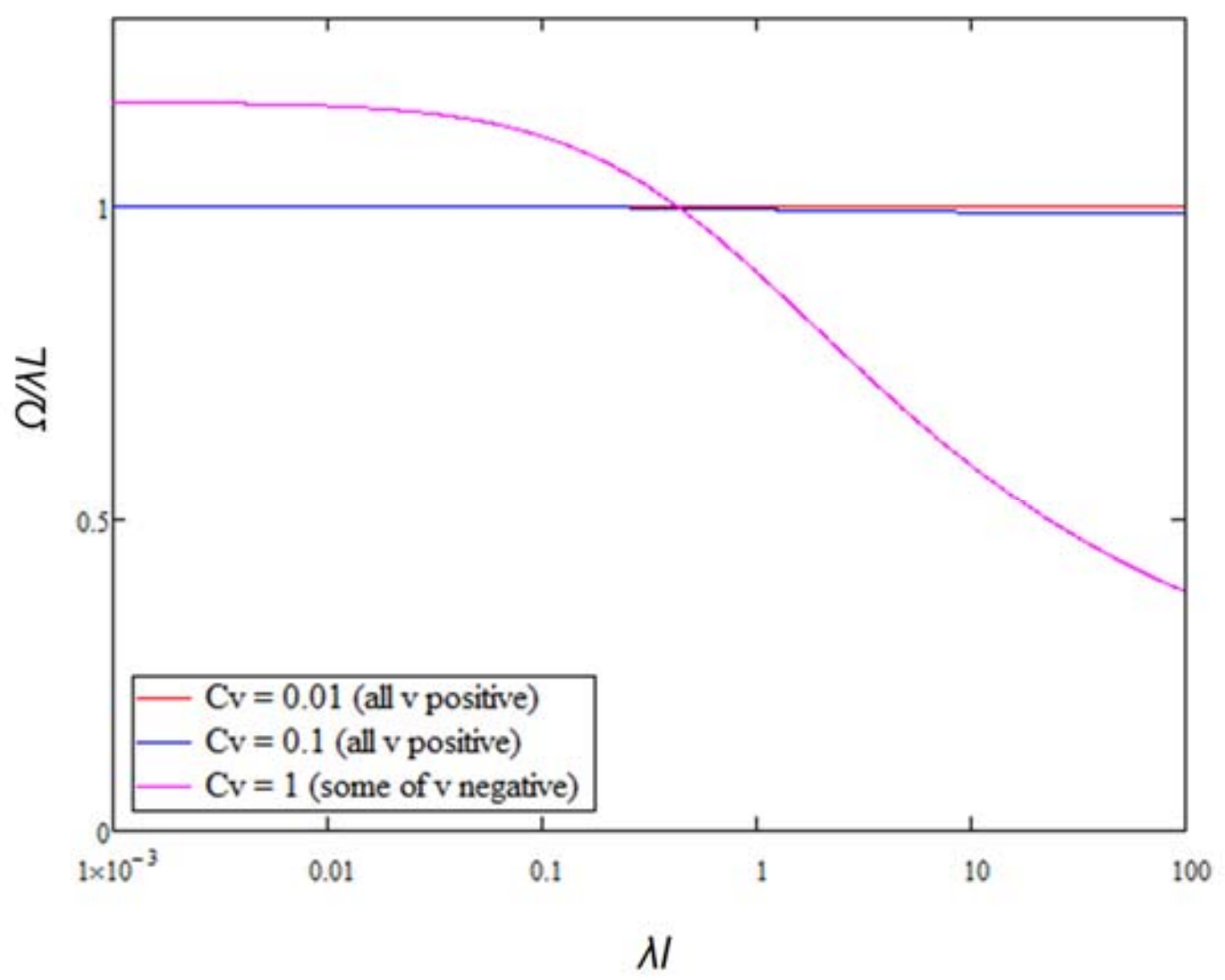

Figure 10. Effect of filtration-mixing number $\lambda l$ on the particle capture ratio $\Omega/\lambda L$.

So, for small and large $\lambda l$, the ratio $\Omega/\lambda L$ tends to a constant, i.e. $\Omega$ becomes proportional to $\lambda L$. At the absence of diffusion ($l$=0), and with only positive velocities, $\Omega$ becomes equal to $\lambda L$.

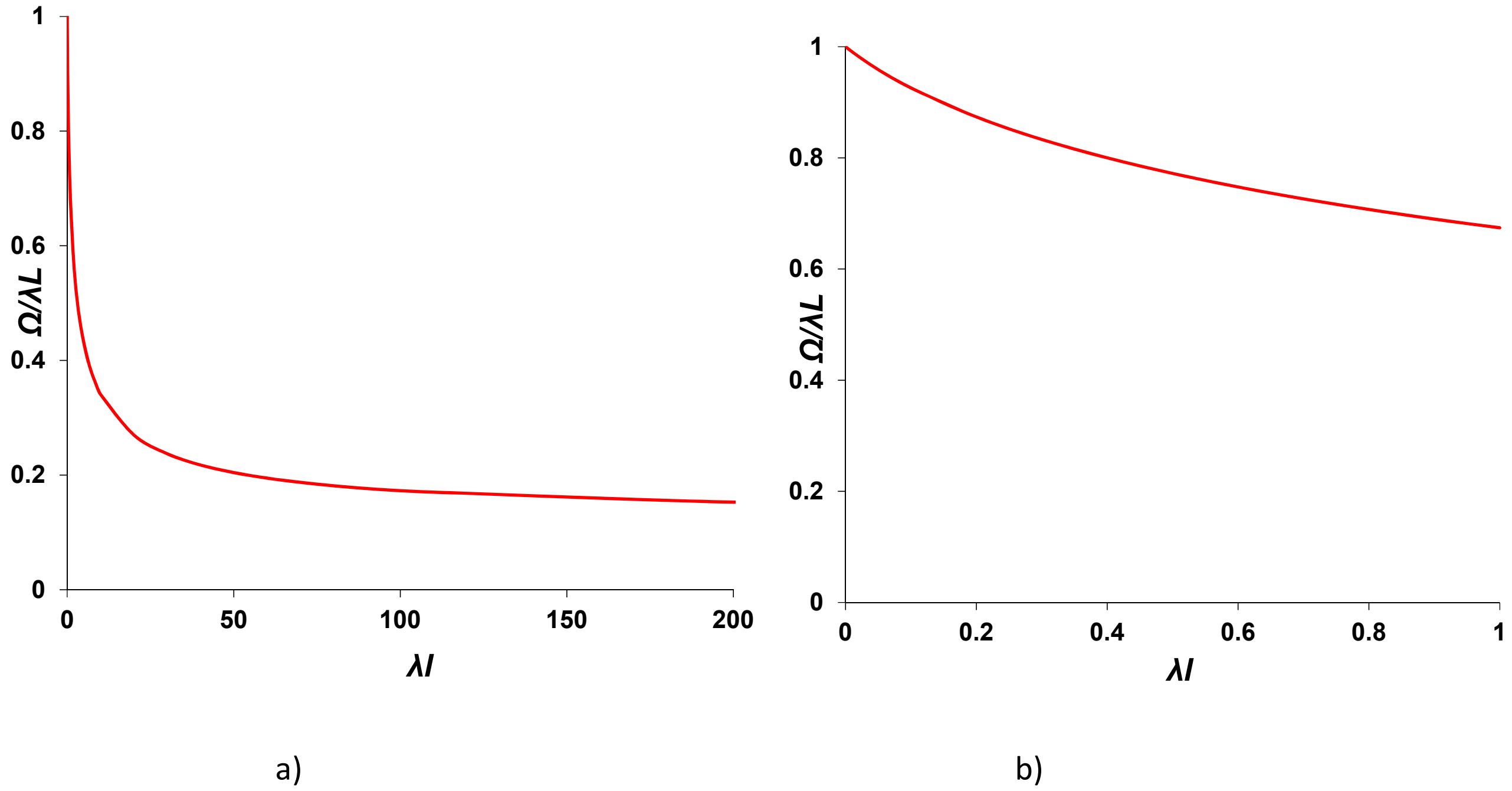

Figure 11. Dependence of dimensionless upscaled filtration coefficient $\Omega/\lambda L$ of dimensionless micro-scale filtration coefficient $\lambda l$: a) overall interval of $\lambda l$; b) zoom.

## 5. Particular cases for advective-dispersive transport with particle capture

In this section we analyse three particular cases of flow — capture-free advective-diffusive transport (section 5.1), advective-free diffusion with capture (section 5.2), and suspension transport with positive particle velocities (section 5.3).

### 5.1. Capture-free suspension flow in porous media

Consider the case of advection with diffusion and without capture. Here model (65) provides the conventional advective-diffusive equation with diffusion coefficient $R_{11}$.

It is proved in Appendix C that $R_{11}$ is proportional to the flow velocity in either the absence or presence of particle capture. This is a well-known fact that follows from different mathematical models and widely observed in laboratory tests [4]

$$R_{11} = \alpha_L \bar{v}\,, \tag{85}$$

where the proportionality coefficient $\alpha_L$ is the dispersivity.

Let us express the dispersivity versus mixing length. The dimensionless Schmidt number, as follows from Eq. (65), in the case of negligible particle capture is

$$\frac{1}{Pe} = \int \frac{y^2}{L/l}\psi_1(y)\,dy - \frac{\left(\int \frac{y}{L/l}\psi_1(y)\,dy\right)^2}{\int \left(L/l\right)^{-1}\psi_1(y)\,dy} = l/L\left[\int y^2\psi_1(y)\,dy - 1\right] = l/L\, C_v^{\,2}. \tag{86}$$

So, the large-scale Schmidt number is equal to the small-scale Schmidt number *l/L* versus the standard deviation of the normalized equilibrium velocity distribution. The ratio of upscaled and small scale Schmidt numbers is equal to the standard deviation of the distribution $\psi_1(y)$.

For the Berea core with the pore size distribution shown in Figure 7a, the standard deviation of the normalised equilibrium velocity distribution $\psi_1(y)$ (Figure 4c) is equal to 0.93; the dispersivity and the mixing length are almost equal. The values of *1/Pe* for zero filtration coefficient (Figure 7b) are almost equal to the dimensionless mixture length *l/L*.

### 5.2. Diffusive advection-free transport with particle capture

Let us discuss the special case with equal probability of particle jumps to the left and to the right

$$\psi_0(-v) = \psi_0(v) \tag{87}$$

i.e. the case of a symmetric equilibrium probability distribution function.

The average velocity in this case equals zero, i.e. it is a pure diffusive case. The transport coefficient $R_{12}$ is also zero, as it is the integral of an even function; see Eq. (B-8).

Calculating the flux and rate using Eqs. (58, 59), one finds that the flux is pure diffusive

$$I = -(B_{11} - B_{00}^{-1}B_1^{\,2})\partial_x c = -R_{11}\partial_x c\,, \tag{88}$$

and the capture rate is

$$\varepsilon = \lambda\left[\overline{|v|} - \lambda\left(B_2 - B_0^{-1}B_{01}^{\,2}\right)\right]c = \lambda\left[\overline{|v|} - \lambda R_{22}\right]c\,. \tag{89}$$

Consider a flux with a uniform initial distribution *c(x,t=0)=c₀*. As follows from (76), the particle distribution remains homogeneous and decays with time with a decrement determined by the particle capture rate:

$$c(X,T) = c_0 e^{-\gamma T},\ \gamma = \lambda|v|\left[1 - \lambda\, R_{22}/|v|\right], \tag{90}$$

A uniform profile is achieved by diffusive Brownian jumps with the same probability in either direction. The upscaled sink term in Eq. (89) and solution (90) describes particle capture by the matrix during these jumps. It results in an exponential decay of the concentration profile with time.

So, for the advection-free diffusivity-dominant flow with $v$=0, the theory (4) results in zero particle capture. The model (5) exhibits the capture that is proportional to the diffusive flux, also resulting in capture-free diffusion. Eq. (89) shows that the upscaled particle capture term is proportional to suspended concentration, which resolves the paradox in the models (4) and (5), mentioned in the Introduction.

### 5.3. Particle flux with positive velocities

Let us consider the case of large advective flow and low diffusion, where the velocities of all particles are positive and equal to their absolute values

$$v = |v| . \tag{91}$$

In particular, the mean velocity is equal to the mean velocity modulus

$$\bar{v} = \overline{|v|} = \int_0^\infty v\psi_0(v)\,dv \ .$$

Substituting Eq. (91) into expressions (B-9)-(B-13) yields

$$B_1 = B_2 = \int_0^\infty (\lambda|v| + \bar{v}/l)^{-1} v\psi_0(v)\,dv , \tag{92}$$

$$B_{11} = B_{12} = \int_0^\infty (\lambda|v| + \bar{v}/l)^{-1} v^2\psi_0(v)\,dv \ . \tag{93}$$

Substituting Eqs. (91-93) into expressions (58, 59) for the flux and capture rate results in

$$q = (\bar{v} - \lambda(B_{11} - B_{00}^{-1}B_1^{\,2}))c - (B_{11} - B_{00}^{-1}B_1^{\,2})\partial_x c , \tag{94}$$

$$\varepsilon = \lambda \ [(\bar{v} - \lambda(B_{11} - B_{00}^{-1}B_1^{\,2}))c - (B_{11} - B_{00}^{-1}B_1^{\,2})\partial_x c] \ . \tag{95}$$

If all velocities are positive, $v=|v|$, then it follows from (B-9)-(B-13) that

$$\varepsilon = \lambda q , \tag{96}$$

which coincides with the capture rate in Eq. (5), and the upscaled filtration coefficient $\Omega/L$ is equal to microscale filtration coefficient $\lambda$. The delay in the mean particle velocity as compared with the carrier water velocity (Eqs. (60, 63)) is equal to $\lambda\alpha_L$, where $\alpha_L$ is the dispersivity coefficient.

Eq. (96) resolves the paradox mentioned in the Introduction [66]. If the Brownian jumps occur in the direction of the advection only, the particle is captured independently on whether it is brought to the retention site by advection or diffusion, and Eq. (5) holds. Otherwise, generalized model (63) describes the particulate transport.

## 6. Discussions

The physical kinetics model of colloidal-suspension-nano transport in porous media includes a particle velocity distribution at each point of the flow domain and the assumptions of diffusive relaxation of the current particle velocity distribution to an equilibrium distribution, and proportionality between the capture rate and particle speed.

*Two modifications for the linearized BGK* version of the Boltzmann's equation are proposed in order to describe flow and capture of colloidal and suspended particles in porous media:

- an arbitrary particle distribution over velocity becomes equal to a given equilibrium distribution $\psi_0(v)$ after particles have travelled a distance $l$ in the porous space;
- the retention is characterized by the probability $\lambda$ for a particle to be captured during its movement over the unit distance (filtration coefficient).

Both constants $l$ and $\lambda$ reflect the difference between the propagation of the particle ensemble in the porous medium and in the "open space".

The introduction of a sink term on the right-hand side of the particle balance equation, instead of non-zero initial data, allows the development of an exact procedure of averaging with exact formulae for transport coefficients of delay, diffusion and capture.

A source term can also be introduced in the right hand side of Boltzmann's equation instead of first type boundary condition. In this case, the source term is proportional to the Dirac delta function $\delta(x-0)$.

*The physical reason for delay* Let us interpret the micro-scale dependencies of macroscale transport coefficients, in particular the expressions (5, 96) for the capture rate. For the geometric model of the porous space of the parallel-pores-and-chambers (Figure 2), the particles are not captured during flow through the pore system. The size-exclusion particle capturing occurs at the chamber outlets. We assume that full fines mixing occurs in the chambers, so the mixing length $l$ is equal to the distance between the chambers. The number of particles crossing the sieve over time $\Delta t$ is equal to $qA\Delta t$, where $A$ is the cross-section area. The $\omega$-th fraction of particles is captured in the sieves, so the number of particles captured by a single sieve is $\omega qA\Delta t$. The volume corresponding to each sieve is equal to $Al$. So, the retention rate is

$$\varepsilon = \frac{\Delta\sigma}{\Delta t} = \frac{\omega qA\Delta t}{Al\Delta t} = \lambda q, \quad \lambda = \frac{\omega}{l} \; . \tag{97}$$

The filtration coefficient is the capture probability per unit length of the particle trajectory; its dimension is $L^{-1}$. Therefore, probability $\omega = \lambda l < 1$. Like in Poisson process, the probability of an event is defined as $\omega = \lambda l$ with $l$ tending to zero; the mean capture-free run is $1/\lambda$. Therefore, only initial sections of dependencies shown in Figures 5-11 have a physical meaning. The delay $\theta$ for $\lambda l < 1$ in Figures 8 and 9 does not exceed one. Otherwise, the particle drift velocity $1-\theta$ is negative, yielding the particle counter flow. In other words, the concentration front in 1D flow propagates contrary to the carrier water, which is physically meaningless.

The above-mentioned speculations of small capture are consistent with the assumption of long waves and large times, used in section 3.2 for the derivation of the constitutive relations (58) and (59).

Figure 3 illustrates size exclusion capture of particles a pore captures a particle if the particle size exceeds the pore size, otherwise the particle passes through the pore. Therefore, the capture rate must be proportional to the total particle flux. A particle is captured by a pore regardless of whether the advective or dispersive flux has brought the particle to the pore, which is the main assumption for the derivation of Eq. (96).

As it follows from Eq. (72), the advective velocity of particles in Eq. (63) differs from the carrier water velocity. This is due to preferential capture of large velocity particles (see the capture law (15) on the micro-scale). So, the capture changes the particle velocity distribution, it removes preferentially fast particles from the flux, which decreases the average particle velocity. Therefore, the advective particle velocity is lower than that of the carrier water, the delay depending on the filtration coefficient.

## 7. Conclusions

The Boltzmann's physical kinetics approach to colloidal-suspension-nano transport in porous media allows drawing the following conclusions.

Substitution of the Cauchy problem with nonzero data by introduction of the sink-source term in the micro-scale Boltzmann's equation and decomposition of the corresponding linear operator in Hilbert space allow for exact upscaling. The upscaled system includes explicit formulae for macroscale model coefficients for dispersion, capture and delay versus micro-scale mixing length, filtration-mixing coefficient, and the variation coefficient of the equilibrium particle velocity distribution.

Under the assumption of fixed mixing length in the microscale Boltzmann's equation, all transport coefficients in the averaged equation are proportional to the mean flow velocity.

The averaged equation differs from the classical deep bed filtration model by the delay in averaged particle velocity as compared with the carrier water velocity, by the diffusion dependence of the filtration coefficient and by the capture rate dependence of the diffusion.

The delay of particle velocity with respect to the carrier water velocity is the distinguished feature of the upscaled system. The delay is explained by preferential capture of fast particles, which decelerates the particle flux. The micro-scale filtration coefficient is defined for small (infinitesimal) trajectory intervals. It explains why the delay does not exceed one, i.e. the particle counter-flow does not occur.

In the absence of capture, the large-scale dispersivity is proportional to the mixing length, and the proportionality coefficient is equal to coefficient of variation for the equilibrium velocity distribution.

For a diffusive advection-free flow with capture, the capture rate is proportional to the filtration coefficient, which is explained by particle capture during their Brownian jumps.

For the particle flux with positive (one-directional) velocities, the capture rate is proportional to the overall particle flux consisting of advective and dispersive (diffusive) components. This corresponds to particle capture by a vacancy independently whether the particle was brought to the vacancy by either advective or diffusive flux. In this case, the upscaled filtration coefficient is equal to its microscale value.

## Appendix A. Derivation of the Operators $L_{ac}$, $L_{aa}$ and $L_{aa}^{-1}$.

Let us first calculate the operator $L_{ac}$ that maps $H_c$ into $H_a$. We apply the operator composition $P_i*()*J_k$ to each of the five components of the operator $L$, (34):

$$
\begin{aligned}
&P_a i\omega J_c = 0 \\
&P_a ikvJ_c = ik\left(v - \langle 1, v\rangle\right) \\
&P_a \lambda_1 |v| J_c = \lambda_1 \left(|v| - \langle 1, |v|\rangle\right). \\
&P_a \frac{\bar{v}}{l}\langle 1, \bullet\rangle J_c = 0 \\
&P_a \frac{\bar{v}}{l} J_c = 0
\end{aligned}
\tag{A-1}
$$

Finally, the explicit expression for the operator $L_{ac}$ is:

$$L_{an} = ik\left[v - \langle 1, v\rangle\right] + \lambda_1\left[|v| - \langle 1, |v|\rangle\right]. \tag{A-2}$$

Now let us calculate the operator $L_{aa}$. Applying the operator $P_a*()*J_a$ to each term of the operator $L$ yields

$$
\begin{aligned}
&P_a i\omega J_a = i\omega \\
&P_a ikvJ_a = ik\left(v\bullet - \langle v, \bullet\rangle\right) \\
&P_a \lambda_1 |v| J_a = \lambda_1 \left(|v|\bullet - \langle |v|, \bullet\rangle\right). \\
&P_a \frac{\bar{v}}{l}\langle 1, \bullet\rangle J_a = 0 \\
&P_a \frac{\bar{v}}{l} J_a = \frac{\bar{v}}{l}
\end{aligned}
\tag{A-3}
$$

The explicit expression for operator $L_{aa}$ is:

$$L_{aa} = i\omega + ik\left(v\bullet - \langle v,\bullet\rangle\right) + \lambda_1\left(|v|\bullet - \langle |v|,\bullet\rangle\right) + \frac{\bar{v}}{l}. \quad \text{(A-4)}$$

The inversion of $L_{aa}$ is realized in solving the following equation:

$$L_{aa}\, p = q, \quad \text{(A-5)}$$

where $p$ and $q$ belong to $H_a$.

$$L_{aa} = P_a(i\omega + ikv + \lambda_1|v| - \frac{\bar{v}}{l}(\langle 1,\bullet\rangle - 1))J_a = P_a(i\omega + ikv + \lambda_1|v| + \frac{\bar{v}}{l})J_a. \quad \text{(A-6)}$$

In order to solve Eq. (A-6), let us rewrite the expression for $L_{aa}$ in the form

$$L_{aa} = P_a \Lambda J_a, \quad \text{(A-7)}$$

where

$$\Lambda = i\omega + ikv + \lambda_1|v| + \frac{\bar{v}}{l}. \quad \text{(A-8)}$$

Now let us solve Eq. (A-5)

$$P_a \Lambda J_a p = q. \quad \text{(A-9)}$$

Since $p$ belongs to $H_a$ and $J_a(p)=p$, from (A-9) it follows that

$$P_a \Lambda p = q. \quad \text{(A-10)}$$

Applying the operator $P_a$ in (A-10) gives

$$\Lambda p - \langle \Lambda, p\rangle = q. \quad \text{(A-11)}$$

Dividing both parts of (A-11) by $\Lambda$ yields

$$p = \Lambda^{-1}\langle \Lambda, p\rangle + \Lambda^{-1} q. \quad \text{(A-12)}$$

Now let us calculate the scalar product of both sides of (A-12) with unit

$$\langle 1, p\rangle = \langle 1, \Lambda^{-1}\rangle\langle \Lambda, p\rangle + \langle \Lambda^{-1}, q\rangle. \quad \text{(A-13)}$$

Since $p$ belongs to $H_a$, the product (A-13) is zero:

$$\langle 1, \Lambda^{-1}\rangle\langle \Lambda, p\rangle + \langle \Lambda^{-1}, q\rangle = 0. \quad \text{(A-14)}$$

This allows calculating the term

$$\langle \Lambda, p\rangle = -\langle 1, \Lambda^{-1}\rangle^{-1}\langle \Lambda^{-1}, q\rangle. \quad \text{(A-15)}$$

Substituting (A-15) into (A-12) yields an explicit solution to Eq. (A-5)

$$p = \Lambda^{-1} q - \Lambda^{-1}\langle 1, \Lambda^{-1}\rangle^{-1}\langle \Lambda^{-1}, q\rangle, \quad \text{(A-16)}$$

and, consequently, to an explicit expression of the inverse operator

$$L_{aa}^{-1} = \Lambda^{-1} - \Lambda^{-1}\left\langle 1, \Lambda^{-1}\right\rangle^{-1}\left\langle \Lambda^{-1}, \bullet\right\rangle . \qquad \text{(A-17)}$$

**Appendix B. Derivation of Constitutive Relations for Long Waves and Large Times**

Let us calculate transport coefficients for the case of long waves and large time scales, where $\omega, k \to 0$.

The operator function of (A-8)

$\Lambda = i\omega + ikv + \lambda|v| + \bar{v}/l$ ,

as $\omega, k \to 0$ becomes

$$\Lambda = \lambda|v| + \bar{v}/l . \qquad \text{(B-1)}$$

Substituting (A-17) into expressions (54) for $R_{ij}$ gives

$$R_{ij} = \left\langle P_a v_i, \Lambda^{-1} P_a v_j - \Lambda^{-1}\frac{\left\langle \Lambda^{-1}, P_a v_j\right\rangle}{\left\langle \Lambda^{-1}, 1\right\rangle}\right\rangle = \left\langle P_a v_i, \Lambda^{-1} P_a v_j\right\rangle - \frac{\left\langle \Lambda^{-1}, P_a v_i\right\rangle\left\langle \Lambda^{-1}, P_a v_j\right\rangle}{\left\langle \Lambda^{-1}, 1\right\rangle} . \qquad \text{(B-2)}$$

Let us calculate the first term on the right-hand side of (B-2), taking into account (B-1),

$$\left\langle P_a v_i, \Lambda^{-1} P_a v_j\right\rangle = \int \left(\lambda_1|v| + \bar{v}/l\right)^{-1}\left(P_a v_i\right)\left(P_a v_j\right)\psi_0 dv = \int \frac{\left(v_i - \left\langle 1, v_i\right\rangle\right)\left(v_j - \left\langle 1, v_j\right\rangle\right)}{\lambda_1|v| + \bar{v}/l}\psi_0 dv =$$
$$= \int \frac{\left(v_i v_j - v_i\left\langle 1, v_j\right\rangle - v_j\left\langle 1, v_i\right\rangle + \left\langle 1, v_i\right\rangle\left\langle 1, v_j\right\rangle\right)}{\lambda_1|v| + \bar{v}/l}\psi_0 dv = \int \frac{v_i v_j}{\lambda_1|v| + \bar{v}/l}\psi_0 dv - \qquad \text{(B-3)}$$
$$-\left\langle 1, v_j\right\rangle \int \frac{v_i}{\lambda_1|v| + \bar{v}/l}\psi_0 dv - \left\langle 1, v_i\right\rangle \int \frac{v_j}{\lambda_1|v| + \bar{v}/l}\psi_0 dv + \left\langle 1, v_i\right\rangle\left\langle 1, v_j\right\rangle \int \left(\lambda_1|v| + + \bar{v}/l\right)^{-1}\psi_0 dv$$

Now calculate the first term in the numerator of the fraction on the right-hand side of (B-2):

$$\left\langle \Lambda^{-1}, P_a v_j\right\rangle = \int \left(\lambda_1|v| + \bar{v}/l\right)^{-1}\left(P_a v_j\right)\psi_0 dv = \int \frac{\left(v_j - \left\langle 1, v_j\right\rangle\right)}{\lambda_1|v| + \bar{v}/l}\psi_0 dv =$$
$$= \int \frac{v_j}{\lambda_1|v| + \bar{v}/l}\psi_0 dv - \left\langle 1, v_j\right\rangle \int \left(\lambda_1|v| + \bar{v}/l\right)^{-1}\psi_0 dv . \qquad \text{(B-4)}$$

Let us now calculate the second term in the numerator of the fraction on the right-hand side of (B-2):

$$\left\langle \Lambda^{-1}, P_a v_i\right\rangle = \int \frac{v_i}{\lambda_1|v| + \bar{v}/l}\psi_0 dv - \left\langle 1, v_i\right\rangle \int \left(\lambda_1|v| + \bar{v}/l\right)^{-1}\psi_0 dv . \qquad \text{(B-5)}$$

The denominator of the fraction on the right-hand side of (B-2) is:

$$\left\langle \Lambda^{-1}, 1\right\rangle = \int \left(\lambda_1|v| + \bar{v}/l\right)^{-1}\psi_0 dv . \qquad \text{(B-6)}$$

Substituting the four terms given by (B-3)-(B-6) into (B-3), one obtains in an explicit expression for the transport coefficients

$$R_{ij}=\int\frac{v_iv_j}{\Lambda}\psi_0dv-\left\langle 1,v_j\right\rangle\int\frac{v_i}{\Lambda}\psi_0dv-\left\langle 1,v_i\right\rangle\int\frac{v_j}{\Lambda}\psi_0dv+\left\langle 1,v_i\right\rangle\left\langle 1,v_j\right\rangle\int\Lambda^{-1}\psi_0dv-$$
$$-\frac{\left(\int\frac{v_j}{\Lambda}\psi_0dv-\left\langle 1,v_j\right\rangle\int\Lambda^{-1}\psi_0dv\right)\left(\int\frac{v_i}{\Lambda}\psi_0dv-\left\langle 1,v_i\right\rangle\int\Lambda^{-1}\psi_0dv\right)}{\int\Lambda^{-1}\psi_0dv}= \tag{B-7}$$

$$=\int\frac{v_iv_j}{\Lambda}\psi_0dv-\left\langle 1,v_j\right\rangle\int\frac{v_i}{\Lambda}\psi_0dv-\left\langle 1,v_i\right\rangle\int\frac{v_j}{\Lambda}\psi_0dv+\left\langle 1,v_i\right\rangle\left\langle 1,v_j\right\rangle\int\Lambda^{-1}\psi_0dv-$$
$$-\frac{\int\frac{v_j}{\Lambda}\psi_0dv\int\frac{v_i}{\Lambda}\psi_0dv-\left\langle 1,v_i\right\rangle\int\frac{v_j}{\Lambda}\psi_0dv\int\Lambda^{-1}\psi_0dv-\left\langle 1,v_j\right\rangle\int\frac{v_i}{\Lambda}\psi_0dv\int\Lambda^{-1}\psi_0dv+\left\langle 1,v_j\right\rangle\left\langle 1,v_i\right\rangle\left(\int\Lambda^{-1}\psi_0dv\right)^2}{\int\Lambda^{-1}\psi_0dv}=$$

$$=\int\frac{v_iv_j}{\Lambda}\psi_0dv-\left\langle 1,v_j\right\rangle\int\frac{v_i}{\Lambda}\psi_0dv-\left\langle 1,v_i\right\rangle\int\frac{v_j}{\Lambda}\psi_0dv+\left\langle 1,v_i\right\rangle\left\langle 1,v_j\right\rangle\int\Lambda^{-1}\psi_0dv-$$
$$-\frac{\int\frac{v_j}{\Lambda}\psi_0dv\int\frac{v_i}{\Lambda}\psi_0dv}{\int\Lambda^{-1}\psi_0dv}+\left\langle 1,v_i\right\rangle\int\frac{v_j}{\Lambda}\psi_0dv+\left\langle 1,v_j\right\rangle\int\frac{v_i}{\Lambda}\psi_0dv-\left\langle 1,v_j\right\rangle\left\langle 1,v_i\right\rangle\int\Lambda^{-1}\psi_0dv$$

Finally, the expression for the transport coefficients reduces to

$$R_{ij}=\int\frac{v_iv_j}{\lambda_1|v|+\bar{v}/l}\psi_0dv-\frac{\left(\int\frac{v_i}{\lambda_1|v|+\bar{v}/l}\psi_0dv\right)\left(\int\frac{v_j}{\lambda_1|v|+\bar{v}/l}\psi_0dv\right)}{\int\left(\lambda_1|v|+\bar{v}/l\right)^{-1}\psi_0dv}. \tag{B-8}$$

Calculate the different terms appearing in (2.11) for *i,j=1,2*.

$$B_{00}=\int(\lambda|v|+\bar{v}/l)^{-1}\psi_0(v)dv=(1,\Lambda^{-1})\ , \tag{B-9}$$

$$B_1=\int(\lambda|v|+\bar{v}/l)^{-1}v\psi_0(v)dv=(v,\Lambda^{-1}1)\,, \tag{B-10}$$

$$B_2=\int(\lambda|v|+\bar{v}/l)^{-1}|v|\psi_0(v)dv=(|v|,\Lambda^{-1}1)\,, \tag{B-11}$$

$$B_{12}=\int(\lambda|v|+\bar{v}/l)^{-1}v|v|\psi_0(v)dv=(|v|,\Lambda^{-1}v)\quad , \tag{B-12}$$

$$B_{11}=\int(\lambda|v|+\bar{v}/l)^{-1}v^2\psi_0(v)dv=(|v|,\Lambda^{-1}|v|)=(v,\Lambda^{-1}v)\,. \tag{B-13}$$

This allows calculating the expressions for the transport coefficients $R_{ij}$:

$$R_{11}=B_{11}-B_{00}^{-1}B_1^2\,, \tag{B-14}$$

$$R_{12}=B_{12}-B_{00}^{-1}B_1B_2\,, \tag{B-15}$$

$$R_{21}=B_{12}-B_{00}^{-1}B_2B_1\,, \tag{B-16}$$

$$R_{22}=B_{11}-B_{00}^{-1}B_2^2\,. \tag{B-17}$$

The expressions of $R_{ij}$ with $\omega,k$=0 are scalars, i.e. they are the same in the space of Fourier transforms and in the real space-time.

**Appendix C. Proof of proportionality between transport coefficients and mean velocity**

Let us prove that the transport coefficients $R_{ij}$ are proportional to the mean velocity $\overline{v}$ .

The equilibrium velocity distribution $\psi_0(v)$ can be normalized by the average velocity $\overline{v}$

$$\psi_0(v)=\frac{1}{\overline{v}}\psi_1(y),\ y=\frac{v}{\overline{v}}, \tag{C-1}$$

where $\psi_1(v)$ is a distribution with unit average.

Substituting (C-1) into expression (B-8) for $R_{ij}$ yields

$$R_{ij}=\overline{v}\int\frac{y_i y_j}{\lambda_1|y|+1/l}\psi_1(y)\,dy-\frac{\overline{v}\left(\int\frac{y_i}{\lambda_1|y|+1/l}\psi_1(y)\,dy\right)\left(\int\frac{y_j}{\lambda_1|y|+1/l}\psi_1(y)\,dy\right)}{\int\left(\lambda_1|y|+1/l\right)^{-1}\psi_1(y)\,dy}\ . \tag{C-2}$$

It is apparent that the transport coefficients $R_{ij}$ are proportional to the mean flow velocity $\overline{v}$ . The ratio between $R_{ij}$ and $\overline{v}$ is equal to the transport coefficient $R_{ij}$ for the normalized velocity distribution. The ratio $R_{ij}/\overline{v}$ is independent of the mean flow velocity $\overline{v}$ .

The mean of the velocity magnitude $\overline{|v|}$ is also proportional to the mean velocity $\overline{v}$

$$\overline{|v|}=\int|v|\psi_0(v)\,dv=\overline{v}\int|y|\psi_1(y)\,dy,\ y=v/\overline{v}, \tag{C-3}$$

and the ratio $\overline{|v|}/\overline{v}$ is independent of the mean flow velocity.

**Acknowledgments**

Many thanks are due to Prof. Anthony Roberts and Dr. Thomas Russell (The University of Adelaide) for fruitful cooperation and meaningful discussions. Many thanks are due to Dr. Themis Carageorgos (University of Adelaide) and Prof. Ch. Alexander (Technical University of Denmark) for their support and encouragement. Dr. A. Zhurov (Russian Academy of Sciences) is gratefully acknowledged for thorough reading the manuscript and improving the text.